\documentclass[twocolumn,tighten]{aastex62}
\pdfoutput=1 
\usepackage{amsmath,amstext}
\usepackage[T1]{fontenc}
\usepackage{apjfonts} 
\usepackage[figure,figure*]{hypcap}
\usepackage{threeparttable}

\newcommand{\aemulus}{{\sc Aemulus }}

\begin{document}
\shorttitle{Aemulus III: Galaxy Clustering}

\title{The Aemulus Project III: Emulation of the Galaxy Correlation Function}


\author{Zhongxu Zhai}
\affiliation{Center for Cosmology and Particle Physics, Department of Physics, New York University, 4 Washington Place, New York, NY 10003, USA}

\author{Jeremy L. Tinker}
\affiliation{Center for Cosmology and Particle Physics, Department of Physics, New York University, 4 Washington Place, New York, NY 10003, USA}

\author{Matthew R. Becker}
\affiliation{Kavli Institute for Particle Astrophysics and Cosmology and Department of Physics, Stanford University, Stanford, CA 94305, USA}
\affiliation{Department of Particle Physics and Astrophysics, SLAC National Accelerator Laboratory, Stanford, CA 94305, USA}
\affiliation{Civis Analytics, Chicago, IL 60607, USA}

\author{Joseph DeRose}
\affiliation{Kavli Institute for Particle Astrophysics and Cosmology and Department of Physics, Stanford University, Stanford, CA 94305, USA}
\affiliation{Department of Particle Physics and Astrophysics, SLAC National Accelerator Laboratory, Stanford, CA 94305, USA}

\author{Yao-Yuan Mao}
\affiliation{Department of Physics and Astronomy and the Pittsburgh Particle Physics, Astrophysics and Cosmology Center (PITT PACC), University of Pittsburgh, Pittsburgh, PA 15260, USA}

\author{Thomas McClintock}
\affiliation{Department of Physics, University of Arizona, Tuscon, AZ 85721, USA}

\author{Sean McLaughlin}
\affiliation{Kavli Institute for Particle Astrophysics and Cosmology and Department of Physics, Stanford University, Stanford, CA 94305, USA}
\affiliation{Department of Particle Physics and Astrophysics, SLAC National Accelerator Laboratory, Stanford, CA 94305, USA}

\author{Eduardo Rozo}
\affiliation{Department of Physics, University of Arizona, Tuscon, AZ 85721, USA}

\author{Risa H. Wechsler}
\affiliation{Kavli Institute for Particle Astrophysics and Cosmology and Department of Physics, Stanford University, Stanford, CA 94305, USA}
\affiliation{Department of Particle Physics and Astrophysics, SLAC National Accelerator Laboratory, Stanford, CA 94305, USA}

\correspondingauthor{Zhongxu~Zhai}
\email{zz681@nyu.edu}

\begin{abstract}

Using the $N$-body simulations of the \aemulus Project, we construct an emulator for the non-linear clustering of galaxies in real and redshift space. We construct our model of galaxy bias using the halo occupation framework, accounting for possible velocity bias. The model includes 15 parameters, including both cosmological and galaxy bias parameters. We demonstrate that our emulator achieves $\sim 1\%$ precision at the scales of interest, $0.1<r<10$ $h^{-1}$Mpc, and recovers the true cosmology when tested against independent simulations. Our primary parameters of interest are related to the growth rate of structure, $f$, and its degenerate combination $f\sigma_8$. Using this emulator, we show that the constraining power on these parameters monotonically increases as smaller scales are included in the analysis, all the way down to 0.1 $h^{-1}$Mpc. For a BOSS-like survey, the constraints on $f\sigma_8$ from $r<30$ $h^{-1}$Mpc scales alone are more than a factor of two tighter than those from the fiducial BOSS analysis of redshift-space clustering using perturbation theory at larger scales. The combination of real- and redshift-space clustering allows us to break the degeneracy between $f$ and $\sigma_8$, yielding a 9\% constraint on $f$ alone for a BOSS-like analysis. The current \aemulus simulations limit this model to surveys of massive galaxies. Future simulations will allow this framework to be extended to all galaxy target types, including emission-line galaxies.

\end{abstract}

\keywords{large-scale structure of universe --- methods: numerical --- methods: statistical}

\section{Introduction}

The spatial distribution of luminous matter is one of the key windows to understanding the distribution and properties of the energy density of the Universe. As galaxy clustering has emerged as an essential tool in our effort to understand the accelerated expansion of the universe, the amount of data from galaxy redshift surveys has increased by orders of magnitude over the past decade. Groundbreaking results from the first truly large-scale redshift surveys, the Sloan Digital Sky Survey (SDSS-I/II, \citealt{SDSS_York, Abazajian_2009}) and the Two Degree Field Galaxy Redshift Survey (2dFGRS, \citealt{Colless_2001, Cole_2005}), have spawned successor programs such as BOSS (\citealt{Dawson_BOSS}), eBOSS (\citealt{eBOSS_Dawson}), WiggleZ (\citealt{Drinkwater_2010}), which are either complete or in progress. Near-term surveys such as the Dark Energy Spectroscopic Instrument (DESI, \citealt{DESI_2016}), 4MOST (\citealt{Jong_2016}), and PFS (\citealt{Takada_2014}) will represent another leap in our ability to create maps of the universe. With the current and near future surveys, we expect to have taken tens of millions of spectra as a community, covering the last 10 billion years of the history of the Universe. In this paper, we propose a new method to make more efficient use of these spectra and increase the constraining power of these datasets. We will demonstrate that incorporating clustering information at nonlinear scales can more than double the power of these data to constrain the growth of structure.

Most of the applications of the these datasets have focused on the retrieval of cosmological information from large scales. The appearance of the baryon acoustic peak at $\sim 100$ $h^{-1}$Mpc allows galaxy clustering to be used as a standard ruler in geometric probes of the expansion history of the Universe, but the detailed shape and amplitude of the measured correlation function also contains significant information. The amplification of clustering through galaxy peculiar velocities---an effect called redshift-space distortions (RSD)---has become the primary method for measuring the growth of structure using spectroscopic surveys. Such measurements are complementary to geometric probes of the Universe because they are less sensitive to variations in the equation of state of dark energy and more sensitive to possible variations in the underlying theory of gravity. Both theories have been proposed to explain the accelerated expansion, but as yet we do not have the data to determine which class of theories is correct.

The design of current and future surveys are built around the rule of thumb $nP\approx 1$, where $n$ is the space density of the targets and $P$ is the amplitude of the power spectrum in the region of interest (\citealt{DESI_2016}). Adhering to this rule creates a survey that is not shot-noise limited, but maximizes volume by limiting the sampling of the density field with a fixed budget of observing time. However, in such survey designs, typical clustering measurements are most precise at the Mpc scale. This is far below the minimum scale considered for analyses of the shape of clustering measurements or the impact of RSD. These analyses are based on variations of higher-order perturbation theory, and usually model clustering at scales down to 30--40 Mpc (e.g., \citealt{Carlson_2009, Carlson_2013}).

Retrieving information from these scales has been a goal of modern cosmology, but it has also been a challenge. Non-linear dynamics of dark matter are captured with excellent precision in modern cosmological $N$-body simulations (see, e.g., \citealt{Klypin_2011, Klypin_2016}). The challenge of constraining cosmology with such simulations is two-fold: (1) an accurate and flexible model of the galaxy bias is required, and (2) one needs to be able to properly sample cosmological parameter space, which becomes computationally intractable for a standard Monte Carlo Markov Chain analysis. Because of these limitations, the amount of information that is extractable from small-scale galaxy clustering is simply unknown. The measurement precision of the data is orders of magnitude higher than at large scales, but the theoretical complexity increases significantly. How much information is recoverable after accounting for all possibilities in galaxy bias? In this paper, for the case of RSD, we will show that after marginalizing over numerous galaxy bias parameters and incorporating the theoretical uncertainty in the galaxy clustering model, it is still possible to extract more constraining power from growth of structure measurements than what is achievable using perturbation theory on large scales.

To solve problem (1), we use the halo occupation distribution (HOD; \citealt{HOD_Weinberg, Peacock_2000, Seljak_2000, Benson_2000, Martin_2001, Cooray_2002}). The HOD approaches galaxy bias by quantifying the statistical relationship between galaxies and dark matter halos. In its most basic form, the HOD is mostly determined by the probability distribution $P(N|M$), the probability that a halo of mass $M$ contains $N$ galaxies of a given class. Once $P(N|M)$ is combined with prescriptions for spatial and velocity bias of galaxies {\it within} halos, this model offers nearly a complete description of the spatial distribution of galaxies for a given halo population. This simple approach of $P(N|M)$, however, ignores the possibility that $N$ may depend on some secondary halo property. If this halo property is correlated with the spatial distribution of halos, this could create a `galaxy secondary bias' (also known as galaxy assembly bias) that would have to be incorporated in the probability distribution in order to create a fully descriptive HOD (\citealt{Sheth_2004, Gao_2005, Harker_2006, Wechsler2006}). The optimal method for incorporating galaxy assembly bias into the HOD, and tests against models that contain these effects, is left to another paper (\citealt{McLaughlin_2018}). Our emphasis here is on determining the total constraining power and the scales from which these constraints come, under the assumption that the assumed HOD approach is sufficient for modeling galaxy bias.

To solve problem (2), we use a combination of novel space-filling algorithms and statistical techniques to create an {\it emulator} for galaxy clustering. Our approach follows from the work of \citet{Heitmann_2009, Heitmann_2010}, \citet{Lawrence_2010}, and \citet{Heitmann_2014}, who created an emulator for the non-linear matter power spectrum. Although full coverage of parameter space with simulations is infeasible, advancements in computing technology and force-calculation algorithms have pushed the field to a state where suites of simulations can be produced and analyzed in a tractable amount of time. If parameter space is properly sampled, novel interpolation schemes can be used to create high-accuracy estimates of statistics at any point within the space. The simulations and the parameter space spanned by the emulator in this paper are described in detail in \citet{DeRose_2018}. These simulations and their application constitute The \aemulus Project. Use of these simulations to emulate the halo mass function is presented in \citet{McClintock_2018}. The goals of the \aemulus Project are not limited to these statistics; these papers are the first step toward a full accounting of the extractable information from small-scale galaxy clustering, including higher-order statistics, void statistics, galaxy--galaxy lensing, and numerous combinations of galaxies with clusters.

Our goal in this paper is to construct an emulator that achieves 1\% accuracy in its prediction of the real-space correlation function and the monopole and quadrupole of the redshift-space correlation function over the scales $1\lesssim r \lesssim 10$ $h^{-1}$Mpc. Motivated both by the mass resolution of the \aemulus simulations and the massive amount of data on massive galaxies, we construct our emulator to model the class of galaxies known as Luminous Red Galaxies (LRGs). Compiling the data from the completed SDSS-I/II and BOSS surveys, as well as the ongoing eBOSS survey, yields nearly two million spectra sampling a volume of nearly 10 Gpc$^3$ (\citealt{eBOSS_Dawson}). This is the ideal target sample to begin the exploration of the constraining power of small-scale clustering. In a pilot study, \citet{Reid_2014} used simulations of a single cosmology to analyze small-scale RSD measurements. Their analysis yielded a constraint on the growth of structure---through the parameter combination $f\sigma_8$---of 2.5\%, a factor of four smaller than analyses using large-scale measurements combined with perturbation theory. In this paper we take the next step forward in this analysis, expanding both the cosmological parameter space as well as the halo occupation parameter space, to produce a robust model with which to analyze massive galaxy datasets.

Our paper is organized as follows. In Section 2, we introduce the method of the emulator and a trial application to the correlation function calculated with analytical methods in real space. Section 3 presents the emulator for the correlation function measured from $N$-body simulations and the performance. Section 4 shows that we can achieve unbiased recovery of cosmological parameters with the emulator, and explores the constraining power of small-scale clustering. We discuss and list our conclusions in Section 5.

\section{Constructing the Emulator}

Here we discuss the observables, the cosmological parameter space, the HOD parameter space, and the implementation of the Gaussian process emulator. We first test this process by building an emulator around an analytical model for one of our observables, the projected two-point galaxy correlation function $w_p(r_p)$. For this statistic, there is a robust analytic model for non-linear galaxy clustering using halo occupation as the galaxy bias model. Here we use the specific implementation described in \cite{Tinker_analytical} and \cite{Tinker_2012}. This model is accurate to 5--10\%, but the accuracy of the model is less relevant than its role as a means by which to test the accuracy of our emulator given the sampling of parameter space, the expected error of our training sample simulations, and the Gaussian process itself.

\subsection{Galaxy clustering: the observables}

The clustering property of the galaxies can be characterized by the two-point correlation function $\xi(r)$, which is defined as a measure of the excess probability, relative to a Poisson distribution, of finding two galaxies at the volume elements $dV_{1}$ and $dV_{2}$ separated by a vector distance $\mathbf{r}$ (\citealt{Peebles1980large}):
\begin{equation}
dP_{12} = n^2[1+\xi(r)]dV_{1}dV_{2},
\end{equation}
where $n$ is the mean number density over the whole sample volume. For a pair of galaxies with redshift-space positions $\mathbf{s}_{1}$ and $\mathbf{s}_{2}$, the dependence of the correlation function is only through $\mathbf{s}=\mathbf{s}_{1}-\mathbf{s}_{2}$ and the orientation of $\mathbf{s}$ relative to the line-of-sight \footnote{We differentiate the correlation function in real space and redshift space as $\xi_{R}$ and $\xi_{Z}$ respectively.}. In this case, we may calculate the correlation function of a two-dimensional grid of separations perpendicular ($r_{p}$) and parallel ($\pi$) to the line of sight $\xi_{Z}(r_{p}, \pi)$ through 
\begin{equation}
\pi = \frac{\mathbf{s}\cdot\mathbf{l}}{|\mathbf{l}|}, \quad r_{p}=\mathbf{s}\cdot\mathbf{s}-\pi^2,
\end{equation}
with $\mathbf{l}=(\mathbf{s}_{1}+\mathbf{s}_{2})/2$ (\citealt{Davis_1983, Fisher_1994}).

In order to mitigate the effect of redshift-space distortions and examine the real-space correlation function, we compute the projected correlation function from $\xi(r_{p}, \pi)$ (\citealt{Davis_1983})
\begin{equation}
w_{p}(r_{p})=2\int_{0}^{\infty}d\pi\xi_{Z}(r_{p}, \pi) = 2\int_{0}^{\infty}d\pi\xi_{R}(r=\sqrt{r_{p}^2+\pi^2}).
\end{equation}
We truncate the integrand to $\pi_{\text{max}}=80$ $h^{-1}$Mpc, which is large enough to include most of the correlated pairs. The projected correlation function eliminates the effect of RSD at all scales $r_{p}<10$ $h^{-1}$Mpc. Coherent inflows can change the amplitude of the $w_{p}$ at larger scales when $\pi_{\text{max}}$ is finite, but in our analytic method we account for this using the linear theory of \cite{Kaiser_1987}. In subsequent calculation based on simulations, any impact of finite $\pi_{\text{max}}$ is already built in to the same result.

Equivalently, we can also write $\xi_{Z}(\mathbf{s}_{1}, \mathbf{s}_{2})$ as $\xi_{Z}(s, \mu)$, where $\mu=r_{p}/s$. We can then expand the correlation function in harmonics of $\mu$ for given $s$, and the resulting Legendre multipoles $\xi_{\ell}(s)$ are given by
\begin{equation}
\xi_{\ell}(s) =\frac{2\ell+1}{2}\int_{-1}^{1} L_{\ell}(\mu)\xi_{Z}(s, \mu)d\mu,
\end{equation}
where $L_{\ell}$ is the Legendre polynomial of order $\ell$.

\subsection{Cosmological Parameter Space}

In our analytic calculation, we apply the spatially flat $\Lambda$CDM model with the following parameters: matter density $\Omega_{m}$, baryon density $\Omega_{b}$, Hubble constant $h\equiv H_{0}(100 \rm{km~s}^{-1} \rm{Mpc}^{-1})^{-1}$, spectral index of the primordial perturbation $n_{s}$, and the perturbation amplitude $\sigma_{8}$. In the simulation-based emulator, we consider flat $w$CDM with the following additional parameters: the constant equation of state of dark energy $w_{0}$ and the number of relativistic species $N_{\rm{eff}}$. We note that these parameters have minimal impact on the small-scale spatial clustering of galaxies. The final parameter used in the simulation-based emulator is $\gamma_{f}$, the amplitude of the halo velocity field relative to the $w$CDM +General Relativity (GR) prediction. The growth rate of structure is the logarithmic derivative of the growth factor, $D$, with $f=d\ln D/d\ln a$. Thus $f$ determines the amplitude of the matter velocity field, which in turn is a product of the matter density and the theory of gravity. Our parameter $\gamma_f$ is defined as $f/f_{w\text{CDM}}$, and all halo velocities are rescaled by this value\footnote{In \cite{Reid_2014}, $\gamma_f$ was labeled $\gamma_{\text{HV}}$.}. $\gamma_f$ is therefore a test of gravity: within the cosmological parameter space allowed by CMB and geometric constraints, can the data be fit with a model where $\gamma_f$ is consistent with unity?

The emulator is based on an efficient parameter sampling strategy and an effective interpolation scheme. The former is realized by a Latin hypercube method as introduced in \cite{Heitmann_2009}, while the latter uses a Gaussian Process (GP). As mentioned in the introduction, we first build the emulator with the analytic method which can provide a thorough estimate of the error and its dependence on scale and location in parameter space. We calculate the matter power spectrum from \cite{EH_1998} and ignore the effect from the equation of state of dark energy and extra relativistic species. This results in a five-dimensional cosmological model with parameters $\Omega_{m}, \Omega_{b}, n_{s}, \sigma_{8}, h$. The 40 training cosmologies for the analytic emulator, as well as 7 test cosmologies, are shown in Figure 3 of \citet{DeRose_2018}. The details of the design of cosmologies can be found from their Table 1. The ranges of the HOD parameters used in this paper are summarized in \autoref{tab:param}. 

\subsection{Halo Occupation Parameter Space}

In this paper, we apply the HOD framework, which approaches the problem of galaxy bias  in a statistical way. In its most basic form, the HOD constructs a probability distribution $P(N|M)$: the probability that a halo of mass $M$ contains $N$ galaxies of a given class. Because the clustering, abundance, and interior structure of dark matter halos is well known from simulations, specifying $P(N|M)$ provides a complete description of the spatial distribution of galaxies. This description is only complete if $N$ depends only on halo mass $M$. If $N$ depends on some secondary halo property, and this halo property depends on the spatial distribution of the halos, this could create a ``galaxy assembly bias" that would have to be incorporated in the HOD model to be fully descriptive. We leave this to another paper (\citealt{McLaughlin_2018}).

For the HOD parameterization, it is necessary to separate the contribution of the central galaxies from that of the satellite galaxies with the mean occupancy of halos:
\begin{equation}\label{N_tot}
\langle N(M)\rangle=\langle N_{\text{gal}}(M) \rangle = \langle N_{\text{cen}}(M) \rangle+\langle N_{\text{sat}}(M) \rangle.
\end{equation} 
The mean number of the central galaxies in each halo is modeled with a smooth transition between 0 and 1 galaxy:
\begin{equation}\label{N_cen}
N_{\text{cen}}(M) = \frac{1}{2} \left[1+\text{erf} \left(\frac{\log_{10}{M}-\log_{10}{M_{\text{min}}}}{\sigma_{\log{M}}}\right)\right],
\end{equation}
and the mean number of satellite galaxies is parameterized as
\begin{equation}\label{N_sat}
N_{\text{sat}}(M) = \left(\frac{M}{M_{\text{sat}}}\right)^{\alpha}\exp{\left(-\frac{M_{\text{cut}}}{M}\right)} N_{\text{cen}}(M).
\end{equation}
The numbers of central and satellite galaxies in the population are drawn from Bernoulli and Poisson distribution respectively. This HOD model has been applied widely in the study of galaxy clustering as first proposed in \cite{Zheng_2005}.
Multiplying the central galaxy occupation function into the satellite occupation function guarantees that the satellite occupation terminates at a mass higher than the central occupation cutoff, i.e. halos cannot host satellites with no central galaxy. In this HOD model, $M_{\text{min}}$, $\sigma_{\log{M}}$, $\alpha$, $M_{\text{sat}}$ and $M_{\text{cut}}$ are the free parameters to be fit by observations which include both $w_{p}(r_{p})$ and the observed number density of galaxies. 
Briefly, $M_{\text{min}}$ is the mass at which half the halos have a central galaxy,
$\sigma_{\log{M}}$ physically relates to the scatter of halo mass at fixed galaxy luminosity, 
$\alpha$ is the power-law index for the mass dependence of the number of satellites, $M_{\text{sat}}$ is a typical mass for halos to host one satellite, and $M_{\text{cut}}$ allows for the cutoff in the satellite occupation function to vary with halo mass.

In the simulation-based emulator, we also introduce three additional parameters related to the halo occupation:

\begin{itemize}
\item $\eta_{\rm{con}}:$ The ratio between the concentration parameters of the satellites and dark matter halos $\eta_{\rm{con}}=c_{\rm{sat}}/c_{\rm{halo}}$, where the dark matter halos are assumed to have Navarro-Frenk-White (NFW) profile (\citealt{NFW_1996}).
   
   \item $\eta_{\rm{vs}}$: The velocity bias parameter for the satellite galaxies, $\sigma_{\rm{sat}}=\eta_{\rm{vs}}\sigma_{\rm{halo}}$, which rescales the velocity of the satellite galaxies relative to the host halos, $\sigma_{\rm{sat}}$ and $\sigma_{\rm{halo}}$ are the velocity dispersion of satellite galaxies and dark matter halos respectively.
   
\item $\eta_{\rm{vc}}$: The same as $\eta_{\rm{vs}}$ but for central galaxies. 
  \end{itemize} 
   
We build a model of galaxy clustering for a sample at $z=0.57$ and a space density of $4.2\times10^{-4}(h^{-1}\text{Mpc})^{-3}$. These choices roughly approximate the BOSS CMASS galaxy sample (\citealt{Reid_2014}). The next paper of this series will apply the model constructed in this work to clustering measurements of CMASS. The techniques developed here are easily applicable to LRGs in current observations such as eBOSS, and near-future observations from the DESI survey. The halo occupation of LRGs has been well studied for a decade, thus we are on firm theoretical ground on which to build our clustering emulator. In this work, we fix the number density of the sample and calculate $M_{\text{min}}$ once all other HOD parameters are known as

\begin{equation}
\bar{n} = \int\frac{dn}{dM} N(M),
\end{equation}
where $dn/dM$ is the halo mass function taken from \cite{Tinker_2008} for the analytic model. In future work, we will use the mass function emulator derived from these simulations. We note that in tests the use of \cite{Tinker_2008} does not bias the results of this paper.

\begin{table*}
\centering
\begin{threeparttable}
\begin{tabular}{llcr}
\hline
& Parameter    & Meaning & Range \\
\hline
Cosmology & $\Omega_{m}$      & The matter energy density    & [0.255, 0.353]   \\
  & $\Omega_{b}$         & The baryon energy density        & [0.039, 0.062]      \\
  &  $\sigma_{8}$ & The amplitude of matter fluctuations on 8 $h^{-1}$Mpc scales.      & [0.575, 0.964]     \\
  &  $h$       & The dimensionless Hubble constant     &  [0.612, 0.748]  \\
  &  $n_{s}$       & The spectral index of the primordial power spectrum     & [0.928, 0.997]    \\
  &  $w$\tnote{$\dagger$} & The dark energy equation of state     & [-1.40, -0.57]   \\
  &  $N_{\text{eff}}$\tnote{$\dagger$} & The number of relativistic species  &  [2.62, 4.28] \\
  &  $\gamma_{f}$\tnote{$\dagger$} & The amplitude of halo velocity field relative to $w$CDM+GR  & [0.5, 1.5] \\
\hline 
  HOD &  $\log{M_{\rm{sat}}}$  & The typical mass scale for halos to host one satellite  & [13.8, 14.5] \\
  &  $\alpha$  & The power-law index for the mass dependence of the number of satellites & [0.2, 1.8]\\
  &  $ \log{M_{\rm{cut}}}$ & The mass cut-off scale for the satellite occupatioin function  &  [10.0, 13.7]\\
  &  $\sigma_{\log{M}}$  & The scatter of halo mass at fixed galaxy luminosity  & [0.05, 0.6]\\
  &  $\eta_{\rm{con}}$\tnote{$\dagger$} & The concentration of satellites relative the dark matter halo   &  [0.2, 2.0]\\
  &  $\eta_{\rm{vc}}$\tnote{$\dagger$}  &  The velocity bias for central galaxies  &   [0.0, 0.7]\\
  &  $\eta_{\rm{vs}}$\tnote{$\dagger$}  & The velocity bias for satellite galaxies  &   [0.2, 2.0]\\
\hline
\end{tabular}
\begin{tablenotes}
\item[$\dagger$] This parameter is not used in the emulator of $w_{p}$ with analytic method.
\end{tablenotes}
\end{threeparttable}
\caption{The parameters used in our emulator, their physical meaning and the range in the parameter space.}
\label{tab:param}
\end{table*}


\subsection{Gaussian Process Parameter Space}

A Gaussian process is a collection of random variables, any finite number of which have a joint Gaussian distribution. For a more detailed discussion of Gaussian processes (GPs) and their features, see \cite{RW_2006}. In the calculation throughout this work, we employ the python code \texttt{george}\footnote{\url{http://george.readthedocs.io/en/latest/}} developed by \cite{george_2014}. 

A Gaussian process can be written as
\begin{equation}
f(\mathbf{x}) \sim \mathcal{GP}(m(\mathbf{x}), k(\mathbf{x}, \mathbf{x'})),
\end{equation}
where $m(\mathbf{x})$ and $k(\mathbf{x}, \mathbf{x'})$ are the mean function and covariance function, respectively. For the sake of simplicity and with no loss of generality, we take the mean function to be zero in the following. Given the input training data $\mathbf{y}=(y_{1}, y_{2}, ..., y_{n})^{T}$ at coordinates $\mathbf{x}=(x_{1}, x_{2}, ..., x_{n})^{T}$ with Gaussian noise $\epsilon\sim\mathcal{N}(0, \sigma_{n}^2)$, we can write the joint distribution of the observation and the function values $\mathbf{f_{\star}}$ at the test locations $\mathbf{x_{\star}}$ as
\[ 
\begin{bmatrix}  \mathbf{y} \\ \mathbf{y_{\star}} 
\end{bmatrix} 
\sim \mathcal{N} \left( 0,  \\ 
\begin{bmatrix} 
K(\mathbf{x}, \mathbf{x})+\sigma_{n}^2 I  & K(\mathbf{x}, \mathbf{x_{\star}}) \\ K(\mathbf{x_{\star}}, \mathbf{x}) &  K(\mathbf{x_{\star}}, \mathbf{x_{\star}}) 
\end{bmatrix} \right),
\]
where $K(\mathbf{x}, \mathbf{x_{\star}})$ denotes the covariance matrix of all the pairs of the training and test points, and the other entries have similar meanings. The conditional distribution of a predicted function value $\mathbf{y_{\star}}$ can then be calculated as
\begin{equation} \label{eq:GP_pre}
\mathbf{y_{\star}}|\mathbf{x}, \mathbf{y}, \mathbf{x_{\star}} \sim \mathcal{N}(\bar{\mathbf{y_{\star}}}, \rm{cov}(\mathbf{y_{\star}})),
\end{equation}
where
\begin{eqnarray}
\bar{\mathbf{y_{\star}}} &= &K(\mathbf{x_{\star}}, \mathbf{x})[K(\mathbf{x}, \mathbf{x})+\sigma_{n}^{2} I]^{-1} \mathbf{y}, \\
\rm{cov}(\mathbf{y_{\star}}) &=& K(\mathbf{x}, \mathbf{x_{\star}})-K(\mathbf{x_{\star}}, \mathbf{x})[K(\mathbf{x}, \mathbf{x})+\sigma_{n}^{2}I]^{-1}K(\mathbf{x}, \mathbf{x_{\star}}). \nonumber
\end{eqnarray}
The remaining problem now is to model the covariance matrix for the training set and test points. This is implemented by choosing a kernel function to populate the elements in the covariance matrix. This is the crucial ingredient in a Gaussian process predictor, as it encodes our assumptions about the function which we wish to learn \citep{RW_2006}. Due to the lack of the knowledge about the correlation in the coordinates, the choice of the kernel function is not restrictive. The basic assumption is that points that are close in parameter space are more strongly correlated than points that are further separated, independent of their absolute location in parameter space. Explicitly, we assume the kernel function in our GP modeling to be a radial basis function which just depends on $r=|\mathbf{x}-\mathbf{x'}|$. Some examples of commonly used kernel functions in literature include the Squared Exponential Covariance Function
\begin{equation}
k_{\text{exp}}(r) = \exp{\Big(-\frac{r^{2}}{2l^{2}}\Big)},
\end{equation}
where the hyperparameter $l$ defines the characteristic length scale. Another example is the Mat\'ern class 
\begin{equation}
k_{\text{Mat\'ern}}(r)=\frac{2^{1-\nu}}{\Gamma(\nu)}\Big(\frac{\sqrt{2\nu}r}{l}\Big)^{\nu}K_{\nu}\Big(\frac{\sqrt{2\nu}r}{l}\Big),
\end{equation}
where $\nu$ and $l$ are the parameters, and $K_{\nu}(r)$ is a modified Bessel function. A special case in machine learning is $\nu=3/2$ which results in
\begin{equation}
k_{3/2} = \Big(1+\frac{\sqrt{3}r}{l}\Big)\exp{\Big(-\frac{\sqrt{3}r}{l}\Big)}.
\end{equation}
In our calculation, we find that a combination of $k_{\text{exp}}(r)+k_{3/2}(r)$ is already flexible enough to model the correlation in the parameter space, and therefore more complicated kernel functions are not used here. The next step is the training of the GP which involves the selection of the hyperparameters of the kernel function. This is through the maximization of the log-likelihood of the training data
\begin{equation}
\ln{\mathcal{L}} = -\frac{1}{2}\mathbf{y}^{T}(K+\sigma_{n}^{2} I)^{-1}\mathbf{y}-\frac{1}{2}\log{|K+\sigma_{n}^{2} I|}-\frac{n}{2}\log{2\pi}.
\end{equation}
After this process is completed and the hyperparameters are known, we can substitute the values into \autoref{eq:GP_pre} to make predictions for the test points.

\subsection{Estimating the Error on the Training Sample}

In order to investigate the performance of the GP in building the emulator, we first apply the above method to calculate $w_{p}$ with the analytical method \citep{Tinker_analytical, Tinker_2012}. We also need to estimate the appropriate error that the emulator will take as input from the training sample. To estimate the error, we use the suite of test simulations. Each of the 7 test cosmologies has 5 realizations (i.e., different initial conditions);  all other simulation properties, including volume and mass resolution, are the same as the training sample. There are two sources of error in the numerical simulations: (1) sample variance in the cosmic structure, and (2) shot noise from the finite number of galaxies per halo. To isolate the sample variance, we take each simulation and populate the halos 10 times with the same HOD but different random seeds. We then take the average value of the clustering over these 10 populations. For each cosmology, we obtain a mean value of the clustering and the 5 deviations from this mean. For the total of 7 cosmologies, we have 35 `differences from the mean,' and the sample variance is taken to be the variance of these 35 numbers. 

For the shot noise, we estimate this from the variance of the 10 populations of the HOD for a single simulation. Thus, we have 35 estimates of the shot noise, from which we take the average. The total error is the quadrature sum of the shot noise and the sample variance, shown in \autoref{fig:wp_analytical_performance}. Sample variance dominates at scales larger than $\sim 1$ $h^{-1}\rm{Mpc}$. When expressed as a fractional error, we find that this error has little dependence on either the cosmology or the HOD parameters. Thus, when estimating the error from the training sample simulations, we apply the result obtained above as a fractional error to all simulations.

\subsection{Implementing the Emulator Using the Analytic Clustering Model}\label{sec:analytic}

\begin{figure}[htbp]
\begin{center}
\includegraphics[width=8cm]{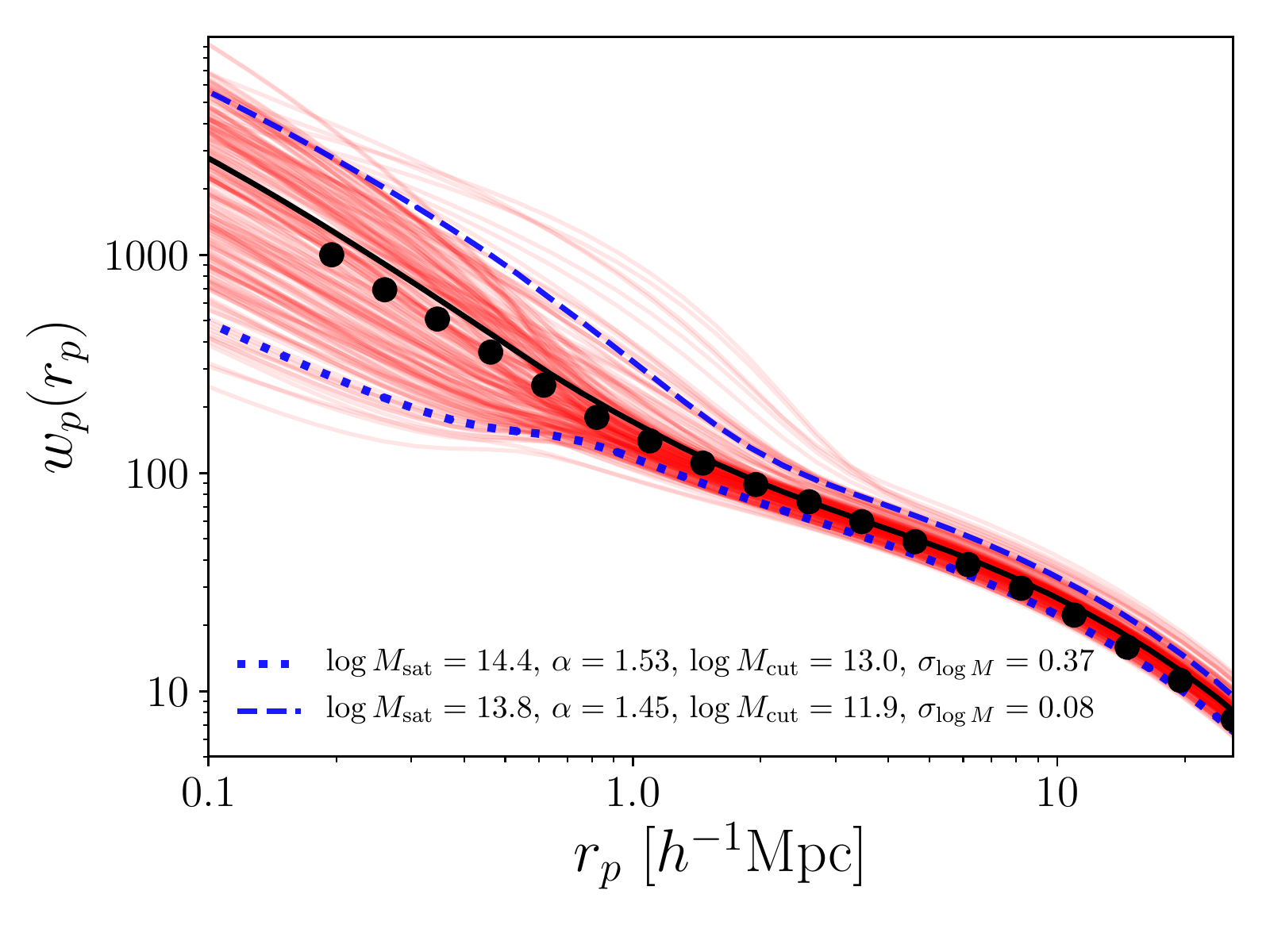}
\caption{A randomly chosen subsample of $w_{p}$ calculated with the analytic method for the input training set (red) and the mean (black). The dotted and dashed blue curves show two HOD models lower and higher than the mean $w_{p}$ respectively with the same cosmology. For comparison, the measurements from BOSS DR11 (\citealt{Reid_2014}) are shown as dots with errorbars ignored for visualization purpose.}
\label{fig:wp_analytical_dis}
\end{center}
\end{figure}

As with the design of the cosmologies, we apply the Latin hypercube method and choose $N_{\text{HOD}}$ designs to sample the HOD parameter space \footnote{For the analytic model, we adopt $N_{\text{HOD}}=400$ and therefore a sampling scheme with overlap; for simulation-based emulator, we choose $N_{\text{HOD}}=2000$ which results a non-overlap sampling scheme when the number of subsample of HOD models is 50.}. \autoref{fig:wp_analytical_dis} shows a randomly chosen subsample of the $w_{p}$ calculated with the analytical method. To reduce memory consumption and CPU time, we select a subsample of HOD models for each cosmology. This provides full coverage of the HOD parameter space without requiring $400\times40$ training points. We construct independent emulators for each bin in $r_{p}$ or $s$ such that the emulator for a specific $r_{p}$ or $s$ bin has its own optimized hyperparameters.  Although this ignores correlation between $r_{p}$ bins, we find that this approach is optimal considering the balance between speed and accuracy. Taking into account the correlation between different $r_{p}$ bins can increase the training set and memory consumption significantly which can affect the optimization of the GP parameters.

\begin{figure}[htbp]
\begin{center}
\includegraphics[width=8cm]{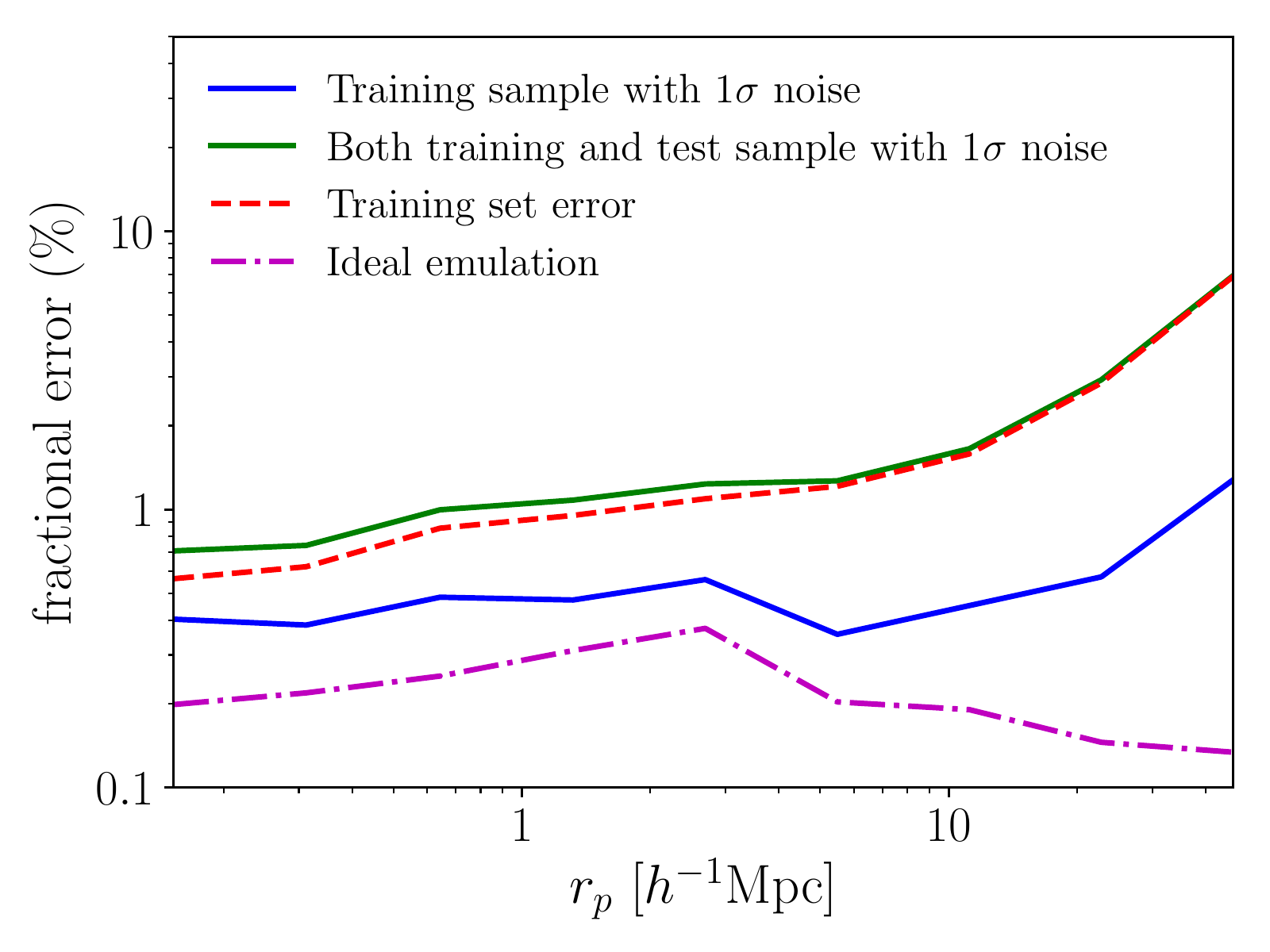}
\caption{The performance of the emulator with the analytic method: the red solid curve is the error level of the input training set ($1\sigma$). The purple line in the bottom represents the emulator performance when the input training sample and test sample have no error. When the training sample is perturbed with $1\sigma$ noise, we retrain the GP and generate emulator predictions to compare with the test sample which has $0$ and $1\sigma$ noise added. The performance is shown by the blue and green lines respectively. The 68\% and RMS error (not shown for clarity) are nearly the same, implying there are no catastrophic outliers. The blue curve represents the error purely due to emulation. Assuming the error from emulation and the error on the test points are independent, the final error estimate of the emulator (green) are the two added in quadrature.}
\label{fig:wp_analytical_performance}
\end{center}
\end{figure}

We first build the emulator under the ideal conditions, where the training data are the original $w_{p}$ calculated with the analytical method and the error $\sigma=0$. In order to evaluate the emulator performance, we first randomly generate 200 cosmology sets and 200 HOD sets within the input parameter space as \autoref{tab:param}, then calculate the $w_{p}$ with the same analytical method at the same scales. We compare the ``truth" with the prediction from the emulator and calculate the fractional error; the result is shown as the purple line in \autoref{fig:wp_analytical_performance}. The error is estimated as the $68\%$ error and root mean square error (the latter is not shown in the figure for clarity) for all the test points, and the results of these two estimates are consistent, implying no catastrophic outliers. This result shows that under the ideal conditions, the emulator can give accurate predictions of the correlation function and that the GP modeling is robust. The error is relatively constant at all scales, with an average value  $0.3\%$.

Next, we build the emulator with the same input training coordinates, but we add a Gaussian random deviation with width of $1\sigma$ to each training data point, where $\sigma$ is the error estimated from the $N$-body simulations as detailed in the previous section. 
The comparison with the test points is shown in \autoref{fig:wp_analytical_performance}. This demonstrates that the accuracy is degraded as we add noise to the input training data, but the decrease is not significant relative to the $1\sigma$ error level itself. Then we add Gaussian noise with a width of $1\sigma$ to the test points as well (green solid line in \autoref{fig:wp_analytical_performance}). The results show that when both the input training data and the test data have noise, the overall error of the emulator is dominated by the error from the test points. This implies that the GP-based emulator can provide estimates of clustering with higher accuracy than the input training sample. We note that this test is constructed over somewhat idealized circumstances; the true error distribution may not be exactly Gaussian, and it assumes that different $r_{p}$ bins are uncorrelated. But even with these caveats, \autoref{fig:wp_analytical_performance} shows the potential of the emulator to make high accuracy predictions of clustering. 

\begin{figure}[htbp]
\begin{center}
\includegraphics[width=\linewidth]{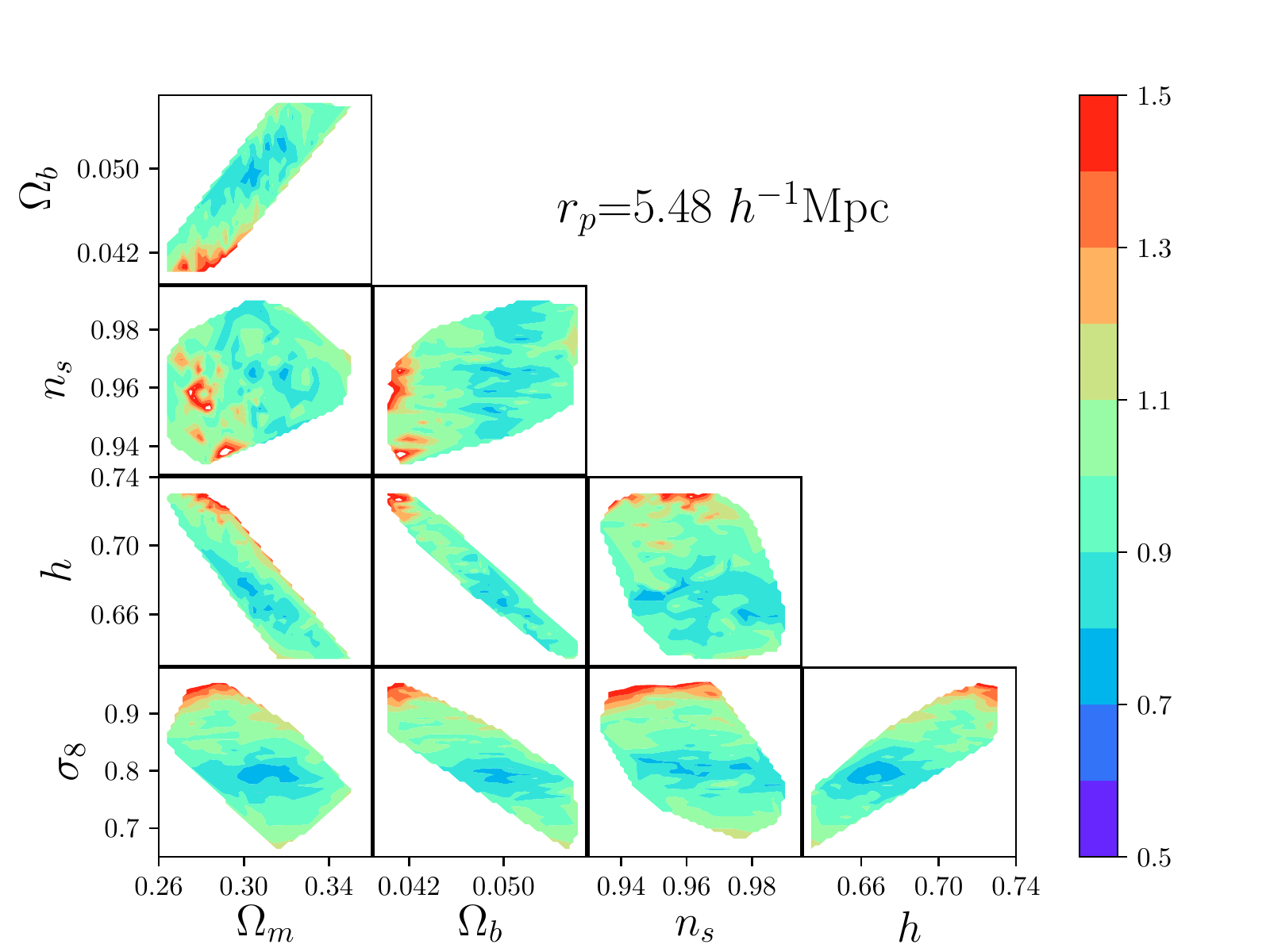}
\caption{The error of the emulator of $w_{p}$ with the analytic method shown in cosmological parameter space, obtained from 200 cosmologies randomly sampled within the training parameter space. The error of each cosmology is estimated from the 68\% error of 200 test HODs. The figure shows error projected onto 2D parameter planes at scale: $r_{p}= 5.48 h^{-1}\rm{Mpc}$, the results for other scales have similar patterns. The errors shown here are normalized by the mean of these errors at this scale (the blue solid line in \autoref{fig:wp_analytical_performance}), and the fluctuations at various positions in the parameter space are mostly around 20-30\%. Compared with the input training error which is a few times larger than the mean, the error from the emulator is fairly constant across the parameter space.}
\label{fig:wp_analytical_error_contour}
\end{center}
\end{figure}

Using the analytic model for $w_{p}$, we can easily explore how the accuracy of the emulator varies with position within the cosmological parameter space. \autoref{fig:wp_analytical_error_contour} shows the $68\%$ error on $w_{p}$ as a contour plot on 2D projections of the cosmological parameter space. The emulator in this test has $1\sigma$ uncertainty on the training sample and the analytic model predictions are taken as truth for the test points (i.e., the blue curve in \autoref{fig:wp_analytical_performance}). The error in this case purely comes from the understanding of the GP from the noisy training set. The result shows that the error of the emulator is more sensitive to the value of $\sigma_{8}$ than other parameters, but the variations of the accuracy are small relative to the $1\sigma$ error level on the training data. This implies that being near the edge in the parameter space does not degrade the emulator accuracy. 

\section{Building the Emulator with $N$-body Simulations}

The previous section demonstrates that the GP is a powerful tool to emulate the galaxy correlation function. We now apply this methodology to the estimate of the correlation function directly from $N$-Body simulations. The details of the simulations are presented in \cite{DeRose_2018}. Briefly, the simulation products are $(1.05~h^{-1} \/\textrm{Gpc})^{3}$ boxes with $1400^3$ particles, resulting in a mass resolution of $3.51\times 10^{10} \left(\frac{\Omega_{m}}{0.3}\right) ~h^{-1}{\rm M}_{\odot}$. Compared with the cosmology designs used for the analytic emulator, the $N$-Body simulations have three extra parameters: $w$, the equation of state of dark energy, $N_{\text{eff}}$, the number of relativistic species, and $\gamma_{f} = f/f_{w\text{CDM}}$, the factor used to scale all halo velocities in the simulation. A fractional change in this parameter is proportional to the change in the linear growth rate at linear and non-linear scales (\citealt{Reid_2014}). In addition, we also add three HOD parameters, $\eta_{\rm{con}}$, $\eta_{\rm{vc}}$, $\eta_{\rm{vs}}$ to incorporate spatial and velocity bias of galaxies within halos.

Our emulator of the galaxy correlation function using $N$-body simulations has 15 parameters in the input parameter set. Because of this increase in dimensionality, we find that a design with 400 HODs is not sufficient to fully sample the space. We increase the number of HOD designs $N_{\text{HOD}}$ to 2000 to obtain the training set. Note that the test cosmologies have five boxes each, so we can get a more accurate estimate of the correlation functions for these test points.

\begin{figure*}
 \center
  \includegraphics[width=0.33\textwidth]{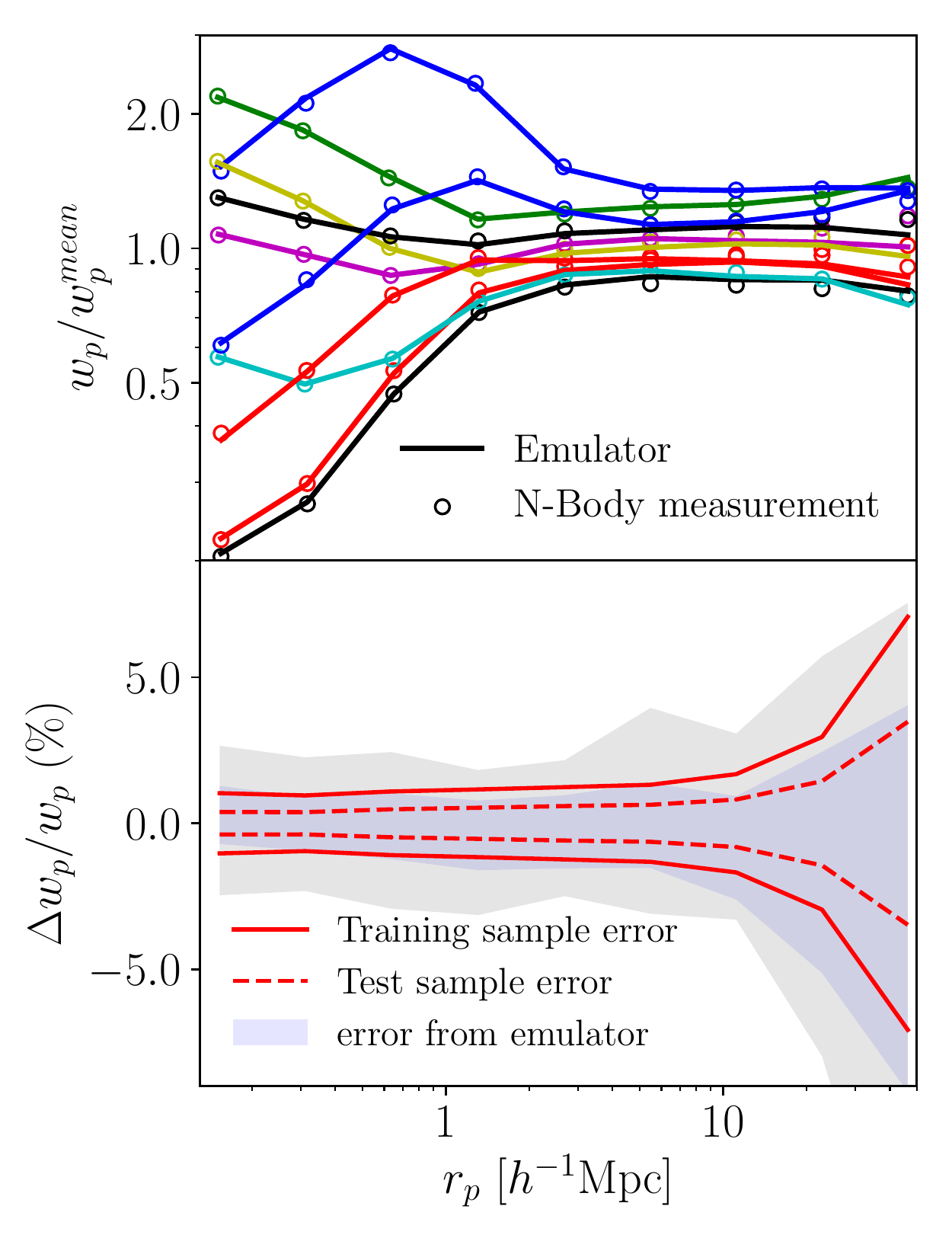} 
  \includegraphics[width=0.33\textwidth]{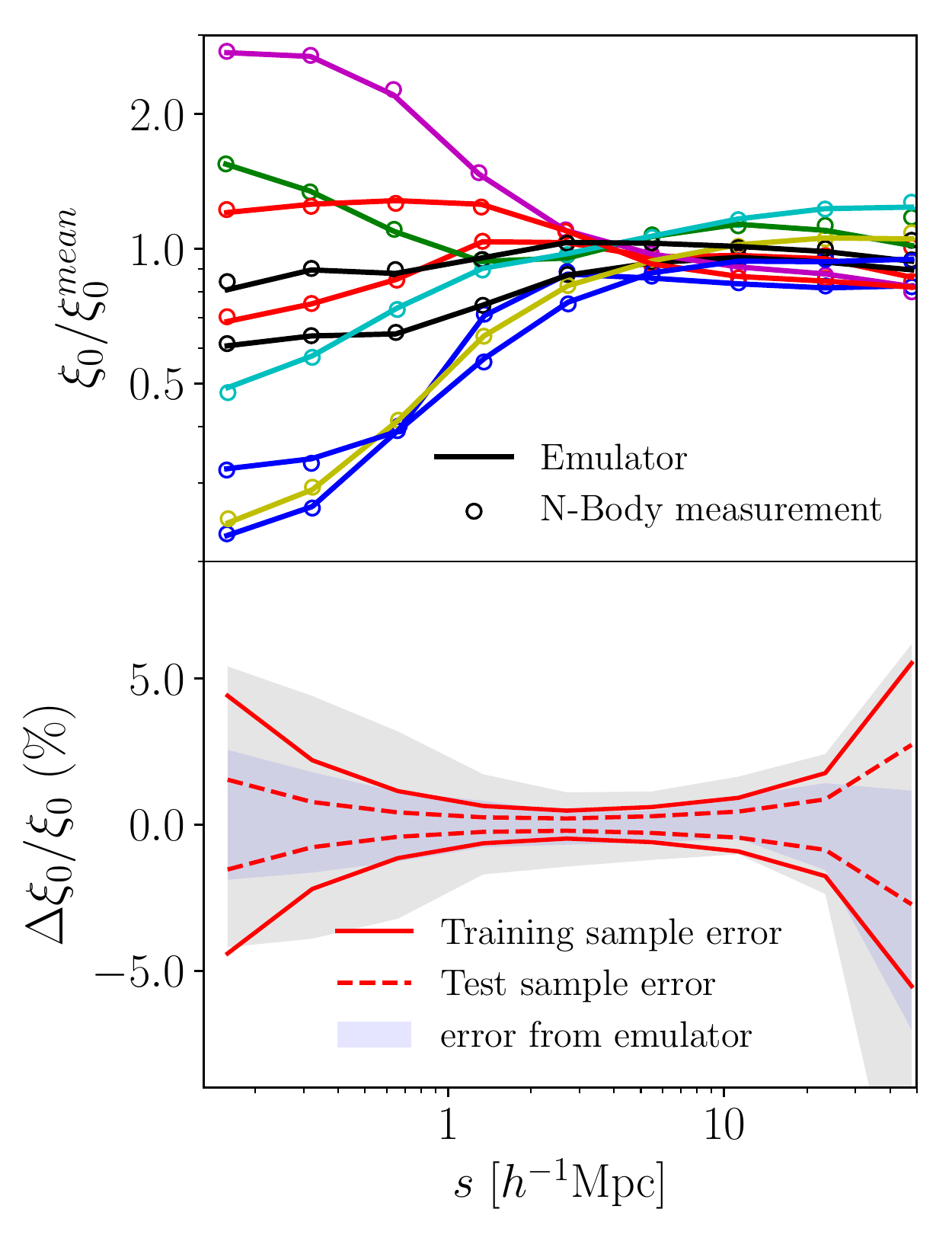}
  \includegraphics[width=0.33\textwidth]{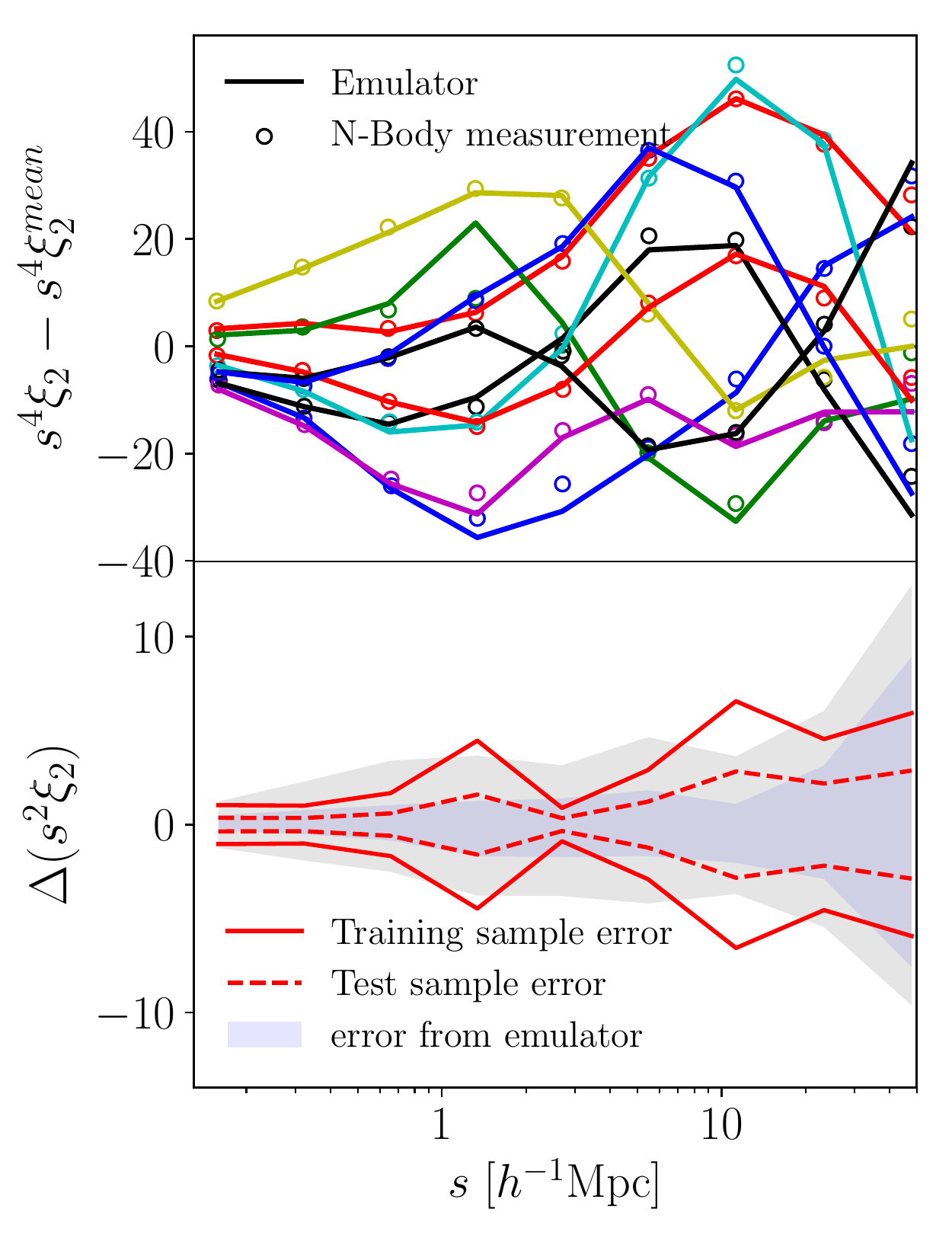}
  \caption{The performance of the emulator for $w_{p}$ (left), $\xi_{0}$ (middle), and $\xi_{2}$ (right) respectively, measured from $N$-body simulations. $Top:$ Ten randomly chosen models for $w_{p}$, $\xi_{0}$ and $\xi_{2}$ measured from simulations and the prediction from emulator. They are shown as the ratio with respect to the mean measurement of the training sample for visualization. $Bottom:$ The 68\% and 95\% error distributions of the emulator, represented by the shaded area. The test sample has error (dashed red) smaller than the training sample (solid red) due to larger simulation volumes. Note that the result for $\xi_{2}$ (right panel) are presented by the absolute error instead of fractional error, and its top panel shows the performance of $\xi_{2}s^{4}$ instead of $\xi_{2}s^{2}$ for plotting purpose.}
  \label{fig:wp_error}
\end{figure*}

We build the emulator for the projected correlation function $w_{p}$, redshift-space monopole $\xi_{0}$, and quadruple $\xi_{2}$ estimated from $N$-body simulations using the same strategy as the previous section for GP modeling, including the kernel functions and scale binning. The accuracy of the emulator is obtained by comparing with the measurements from test boxes which contain seven cosmologies, randomly choosing 100 HOD sets in the same parameters space as the input training sample. This results a test sample of 700 models. The three columns of \autoref{fig:wp_error} presents the performance of the emulator, for $w_{p}$, $\xi_{0}$ 
, and $\xi_{2}$ respectively.

The top left panel of \autoref{fig:wp_error} shows a few examples of $w_{p}$ chosen to spread over the $w_{p}$ amplitude calculated from the emulator and directly from $N$-body simulations respectively. It shows that the emulator can generate high-accuracy predictions. The bottom left panel shows the performance of $w_{p}$ from the emulator. The solid red line is the error of the input training set which has single population of the dark matter halos for each position in the parameter space. The dashed red line shows the error for the test points which is smaller than the training error, since the test boxes have larger volume with multiple populations to suppress the shot noise and sample variance. The emulator error is represented by the distribution of the shaded area (1 and $2\sigma$ respectively). The overall error of the emulator is 0.9\% in the $1-10 h^{-1}\rm{Mpc}$ range. 

The middle column of \autoref{fig:wp_error} shows a similar result for the monopole $\xi_{0}(s)$. A similar conclusion can be drawn about this emulator. Note that the errors for the training sample and the emulator is larger than $w_{p}$ at small scales due to the shot noise. The right column of \autoref{fig:wp_error} represents the result for quadrupole $\xi_{2}(s)$. Because $\xi_{2}(s)$ approaches zero and goes negative at some scale, the fractional error is not a useful statistic for the emulator performance; here the results from the emulator and the test boxes are shown as absolute errors. The overall $1\sigma$ error of the emulator is smaller than 0.7 of the training error at the scales of interest. 

The error as estimated from the test points is smaller than the error on the training sample, implying that the emulator is performing better than the errors on its inputs. This demonstrates that as long as the error of the input training set is smaller than 1\%, the emulator can generate predictions better than 1\%. In the following calculation, we define the pure error from the emulator as the emulator uncertainty, which is assumed to be independent from the error of the test points. The addition of these error budgets in quadrature gives the total error of the emulation represented by the envelope of the shaded area in the bottom panels of \autoref{fig:wp_error}. Compared with the $w_{p}$ emulator using the analytic method in \autoref{sec:analytic}, the simulation-based emulator does not perform much better than the training error. This is partly due to the extended parameter space, and perhaps more importantly to the ignorance of the correlation between different statistics and $r_{p}$ or $s$ bins. Taking into account this correlation will increase the training sample for individual emulators and will increase memory consumption, which implies that other interpolating schemes beyond GP is worth investigating in the future work.

\section{Recovery of cosmological parameters and constraint from the emulator}

\begin{figure*}
 \center
  \includegraphics[width=0.32\textwidth]{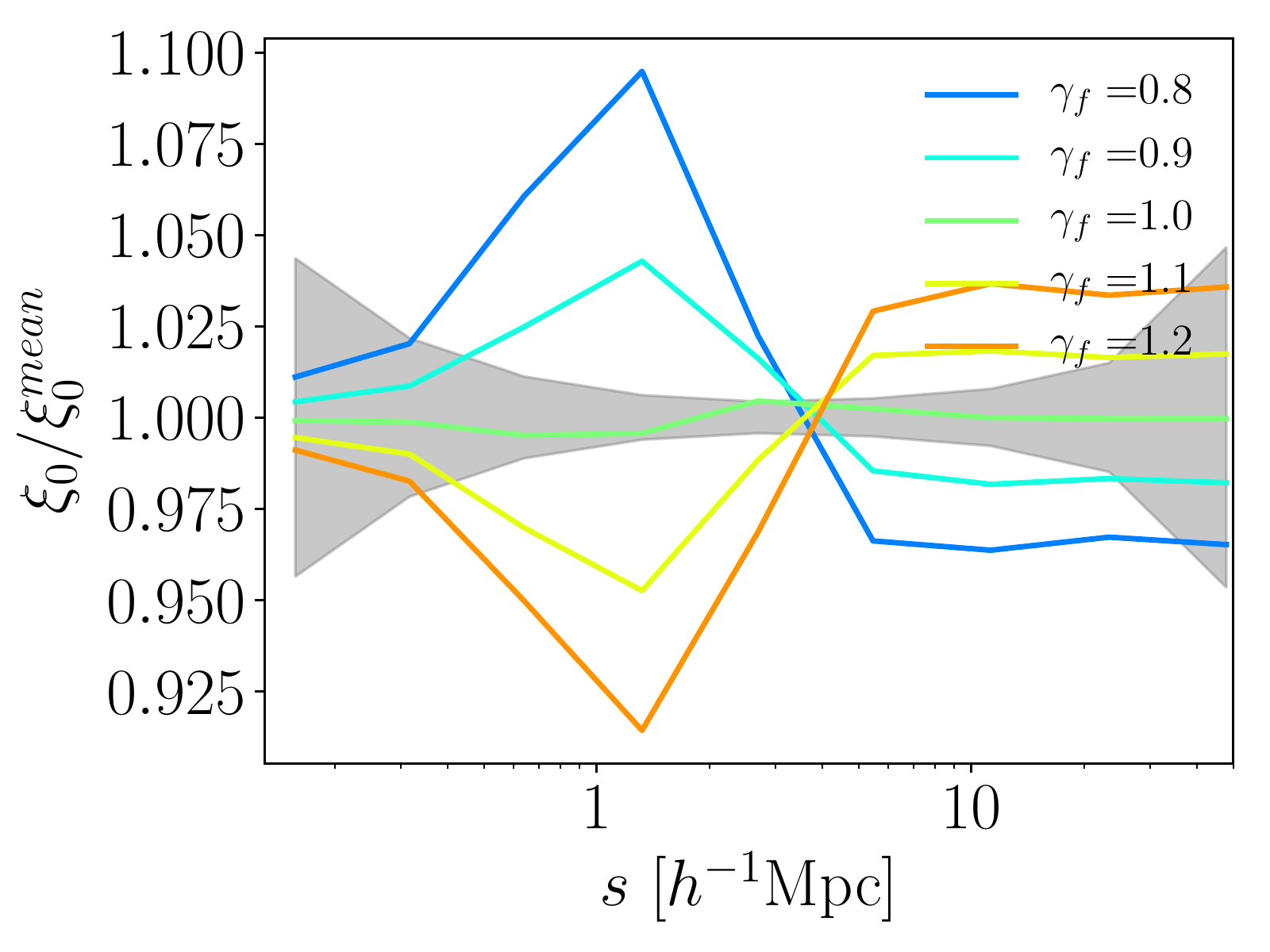}
  \includegraphics[width=0.32\textwidth]{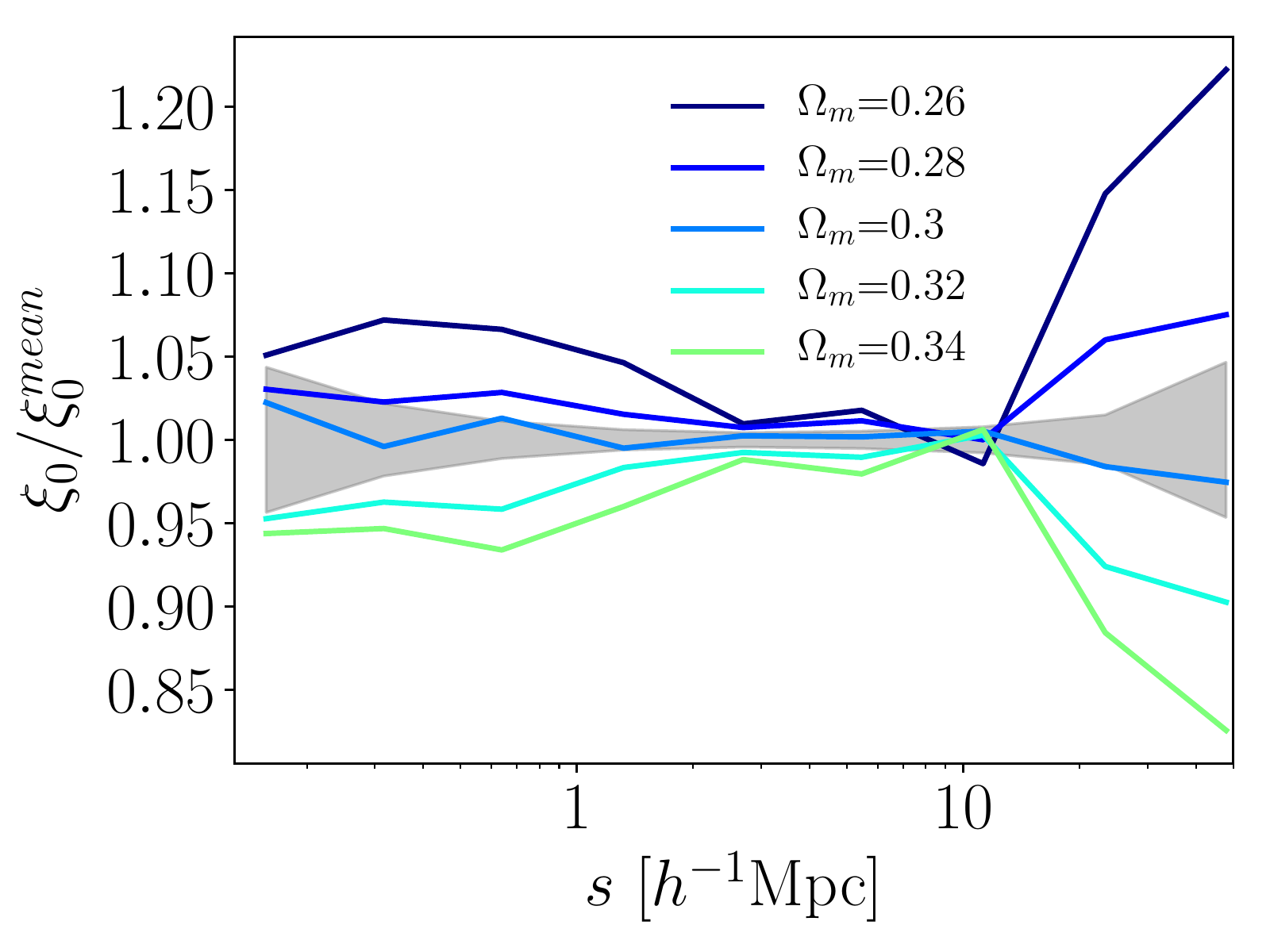}
    \includegraphics[width=0.32\textwidth]{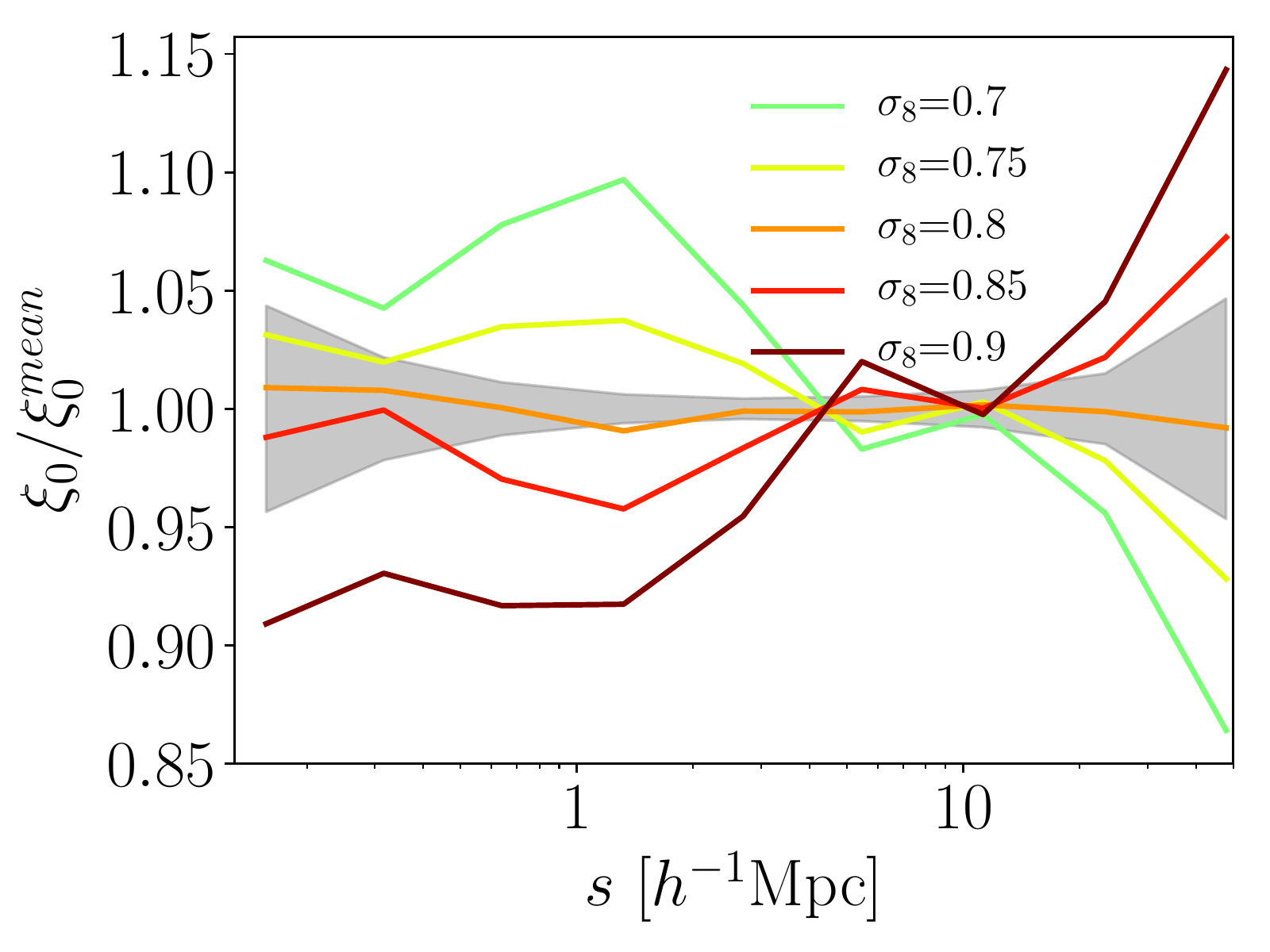}
  \caption{The cosmological parameter dependence of the redshift-space monopole given the real-space clustering amplitude. In the test, we fix all the cosmological parameters and perform a HOD fit to the given $w_{p}$. From left to right, the cosmological parameters are the same except $\gamma_{f}$ (left), $\Omega_{m}$ (middle) and $\sigma_{8}$ (right) as labeled. The grey band shows the sample variance equivalent to the BOSS DR11 volume.}
  \label{fig:test_f_Om_s8}
\end{figure*}

In linear theory, redshift-space distortions carry information about coherent flows of matter into overdense region. This boosts the amplitudes of the RSD multipoles in a manner that is only dependent on the degenerate parameter combination $f\sigma_8$, where $f=\gamma_{f}f_{w\text{CDM}}$, and $f_{w\text{CDM}}$ is the linear growth rate in the cold dark matter cosmology. Scaling this factor by the halo velocity parameter $\gamma_{f}$ gives a direct estimate of the linear growth of structure \citep{Reid_2014}. At translinear and non-linear scales, the impact of peculiar velocities is more than simply a change in amplitude. Non-linearities are most obviously present in the so-called finger-of-god effect, seen as an elongation of contours of the correlation function along the line of sight (see, e.g., \citealt{Peacock_2000}). The amplitude of this effect is sensitive to the mass scale of the halos themselves, which is sensitive to the abundance of massive halos and to the details of the galaxy bias model. 

In \autoref{fig:test_f_Om_s8} we present a pedagogical test in which we vary a single cosmological parameter and show the impact on the RSD monopole. Each time a parameter is varied, the real-space correlation function, $w_p(r_p)$ is re-fit to determine new HOD parameters. Only HOD parameters that control the mean occupation function are allowed to vary. In the left-hand panel, we vary the amplitude of the halo velocity field, $\gamma_{f}$, from 0.8 to 1.2. The real-space clustering of each model is nearly identical in each fit. At $s>5$ $h^{-1}$Mpc, we have the expected linear behavior where increasing $f$ increases the amplitude of $\xi_0(s)$. At smaller scales, however, increasing $f$ suppresses the monopole. This is due to the increase in random motions of close halo pairs, spreading these pairs out along the line of sight. This suppression goes away at the smallest scales, where the pairs are mostly within a single halo. The middle panel of \autoref{fig:test_f_Om_s8} shows variations in $\Omega_m$. Although increasing the matter density increases the amplitude of the velocity field akin to changing $f$, in this panel we do not see the expected linear behavior at large scales. This is because of the impact $\Omega_m$ has on the shape of the matter correlation function; increasing $\Omega_m$ at fixed $\sigma_8$ increases large-scale power at $s>10$ $h^{-1}$Mpc. In each of these models, the fit to $w_p$ is consistent at $r_p<10$ $h^{-1}$Mpc, but at larger scales the galaxy correlation functions diverge, leading to the divergence seen in $\xi_0(s)$. At $s<10$ $h^{-1}$Mpc, non-linearities are already dominant over coherent velocity flows, producing a trend of lower $\Omega_m$ yielding higher $\xi_0(s)$. Changing the matter density also changes the mass of dark matter halos, thus galaxy pairs within halos therefore have higher relative velocities. The right-hand panel shows variations in $\sigma_8$ at $z=0$. Increasing $\sigma_8$ increases the large-scale power in redshift space, as expected. But $\sigma_8$ also influences the abundance of massive halos: larger $\sigma_8$ results in more cluster-sized objects, which are the source of galaxy pairs within the same halo at $r\approx 1$ $h^{-1}$Mpc. Because the halos are massive, the relative velocities of galaxies within these halos is large, leading to increased suppression of redshift-space clustering at the transition between one-halo and two-halo clustering. 

These parameters impact galaxy velocities at large and small scales, as well as the shape of real-space clustering. The combination of real- and redshift-space observables therefore provides enough leverage to break degeneracies and independently constrain each of these three quantities, without the need for a strong CMB prior. In this section, we demonstrate the ability of our clustering emulator to recover the input cosmologies on both the test $N$-body simulations, and with even higher resolution simulations that can track substructure within halos. 

\subsection{Recovery test on simulation boxes} 
\label{sec:recov}

\begin{figure*}
 \center
  \includegraphics[width=1.0\textwidth]{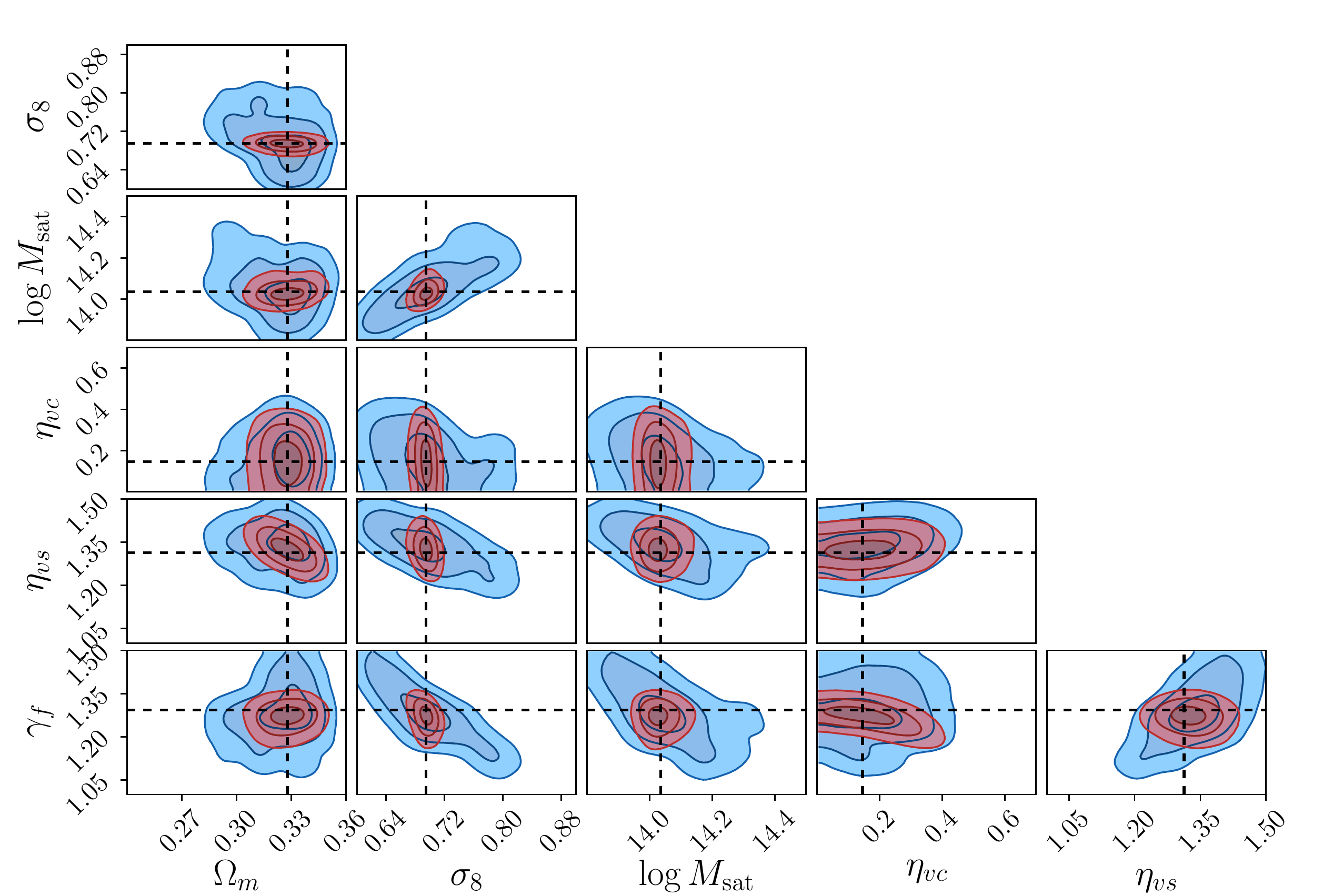}
  \caption{Recovery test with the emulator on a randomly chosen test cosmology. The blue contours show the 1- and 2-$\sigma$ constraint on a subset of the cosmological and HOD parameters based on the correlation function measurements. The red contours indicate results with a Planck prior. The true (input of the test cosmology) parameters are marked as dashed lines.}
  \label{fig:con_testBOX}
\end{figure*}

\begin{figure*}
 \center
  \includegraphics[width=0.45\textwidth]{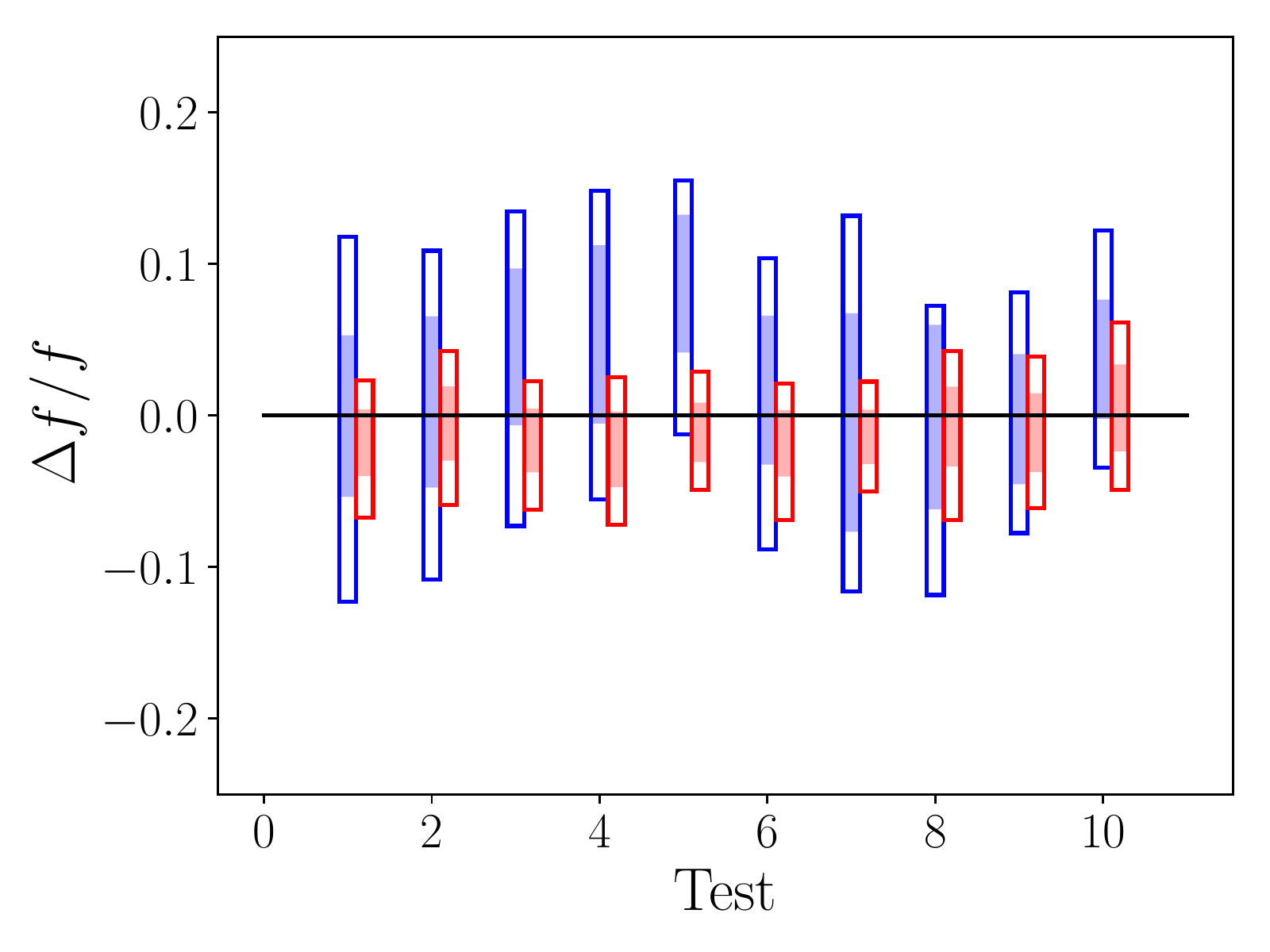}
  \includegraphics[width=0.45\textwidth]{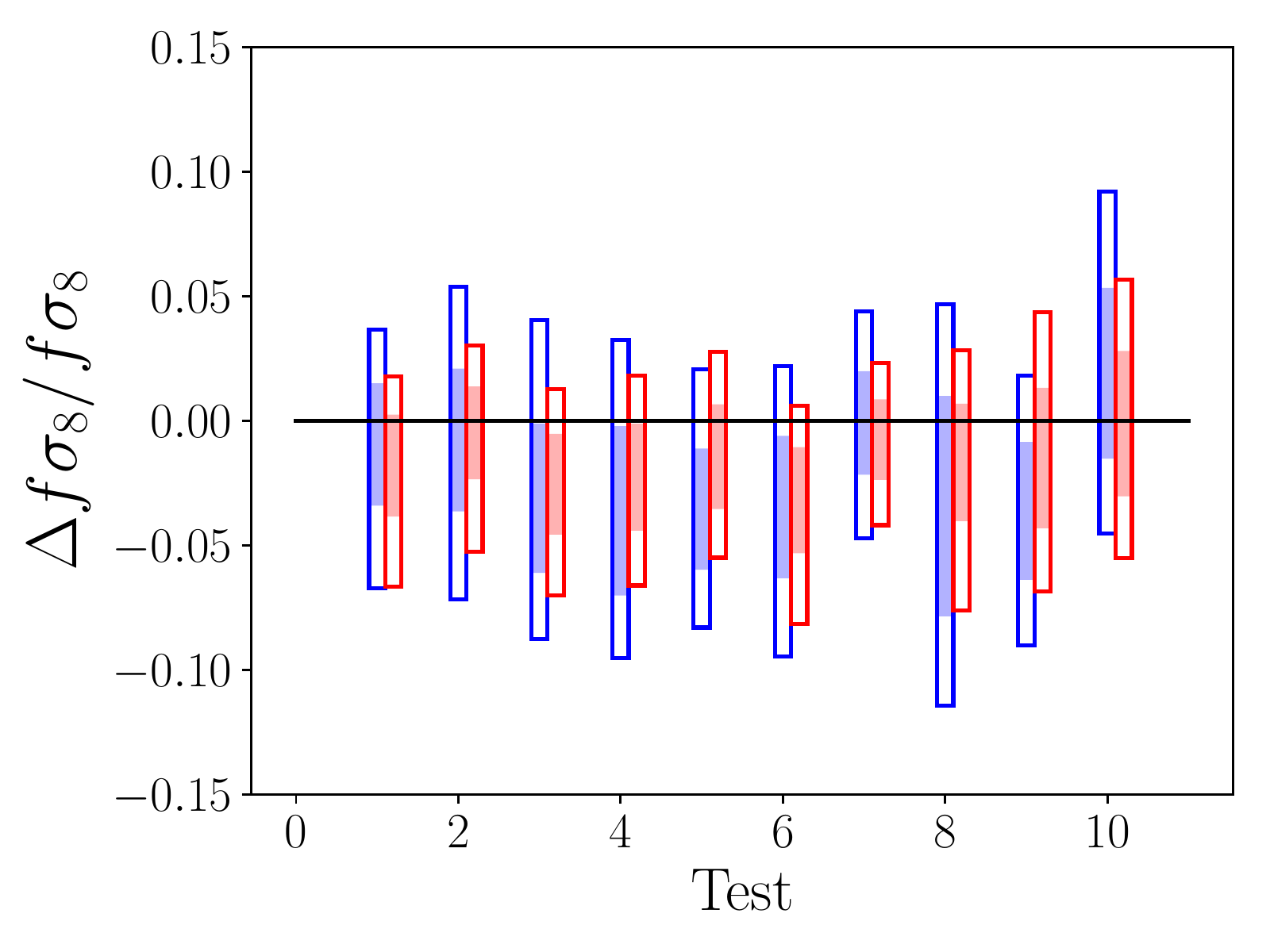}
  \caption{The constraints on $f$ (left) and $f\sigma_{8}$ (right) for 10 randomly chosen cosmology+HOD models, shown as fractional error. The filled and open squares show 1 and $2\sigma$ constraint respectively. The blue and red bands show results without and with Planck priors. The average 1(2)$\sigma$ constraint on $f$ from all these models is 5.3(9.9)\% without and 2.3(4.7)\% with Planck priors respectively, while the corresponding results for $f\sigma_{8}$ are 3.0(6.0)\% without and 2.2(4.5)\% with Planck priors.}
  \label{fig:con_f_fsigma8}
\end{figure*}

With the above emulator for galaxy correlation function, we can now fit a given measurement in cosmological and galaxy occupation parameter space. The result can also provide a direct measurement of $f\sigma_{8}$ at $z\sim0.57$. Moreover, the combination of real- and redshift-space clustering affords enough constraining power to break this degeneracy and constrain $f$ directly. As a first test, we randomly choose a parameter set (including both cosmological parameters and HOD parameters) from the test boxes, and assume that the correlation function measurements are ``observational data''. Then we use our emulator to generate predictions and construct a likelihood function. 
To estimate the likelihood of a given model, we need to estimate the covariance in the data. This includes not just the correlation between $r$-bins of a given statistic, but also between the statistics themselves. In order to estimate this correlation between $w_{p}$, $\xi_{0}$, and $\xi_{2}$ for the likelihood function, we use the Minerva simulations, a set of 100 $N$-body simulations \citep{Grieb_2016}. The parameters of the HOD model are chosen to be ``CMASS-like'' at $z\sim0.57$, but in our test we find that the errors depend little on the HOD model. We calculate the galaxy correlation function from these galaxy catalogs and estimate the correlation matrix (by normalizing the covariance matrix for these galaxy correlation functions).  As a conservative test, we first choose the error of the correlation function to be an addition in quadrature of the input training error of the emulator (which corresponds to a simulation volume of about 5 ($h^{-1}$Gpc)$^{3}$) and the emulator uncertainty as defined above. The resulting error thus corresponds to a simulation volume smaller than 5 ($h^{-1}$Gpc)$^{3}$. We use this error estimate as the diagonal elements combined with the correlation matrix to populate the covariance matrix in our construction of the likelihood function.\footnote{This implicitly assumes that the emulator prediction has identical correlation as the ``observation''. This is not guaranteed in the application to real data, but the final result shouldn't be affected significantly with a more realistic model.}

The resulting likelihood function can be written as
\begin{equation}\label{eq:likelihood}
\ln{\mathcal{L}} = -\frac{1}{2}(\xi_{\text{emu}}-\xi_{\text{obs}})C^{-1}(\xi_{\text{emu}}-\xi_{\text{obs}}),
\end{equation}
where $\xi_{\text{emu}}$ and $\xi_{\text{obs}}$ are the correlation function from the emulator and observation respectively and $C$ is the covariance matrix as defined above. The exploration is obtained through an MCMC test with the \textsc{python} package \texttt{emcee}\footnote{\url{http://dfm.io/emcee/current/}}(\citealt{ForemanMackey_2013}) which is based on an affine-invariant ensemble sampling algorithm (\citealt{GW_2010}). \autoref{fig:con_testBOX} shows constraints on the key cosmological parameters of interest, as well as key HOD parameters that could show degeneracies with cosmology. This result is obtained with flat priors on the parameters, which are uninformative. As expected, there is a strong degeneracy between $\gamma_{f}$ and $\sigma_{8}$. The HOD parameter most degenerate with $\gamma_{f}$ is $M_{\rm{sat}}$; lowering $M_{\rm{sat}}$ increases the bias of the sample, reducing the amplitude of the two-halo redshift-space clustering. To counterbalance this, $\gamma_{f}$ increases the redshift distortions. This is the same reason for the degeneracy between $\gamma_{f}$ and $\sigma_{8}$ --- lower $\sigma_{8}$ requires a higher $M_{\rm{sat}}$ and bias to fit $w_{p}$. For reference, the true cosmology is indicated with the black cross.

The above constraints are obtained with no CMB priors other than flat priors that define the parameter space of our simulations. We also apply the constraint from Planck measurement as priors of the cosmological parameter. In particular, we choose the constraint on $\Omega_{m}$, $\Omega_{b}$, $h$, $\sigma_{8}$ and $n_{s}$ from \texttt{TT+lowP+lensing} chain (\citealt{Planck_2015}) of the $\Lambda$CDM model. The constraints are shown as the red contours in \autoref{fig:con_testBOX}. Applying this Planck prior significantly strengthens the constraints on both cosmology and HOD parameters. This is primarily due to the strong prior on $\sigma_{8}$ introduced by the Planck data.

We repeat the above test for 10 randomly chosen test simulation boxes and HODs. The fractional errors on $f$ and $f\sigma_{8}$ are shown in \autoref{fig:con_f_fsigma8}. The fractional error of $f$ is mostly at 5\% or higher, and can be improved significantly by the Planck data to 2--3\%. However, for the product of $f\sigma_{8}$, the improvement from Planck data is marginal compared with correlation function only which constrains $f\sigma_{8}$ to 2--3\% level.

\subsection{Scale-dependence of the constraint on structure growth}

\begin{figure*}
 \center
  \includegraphics[width=0.45\textwidth]{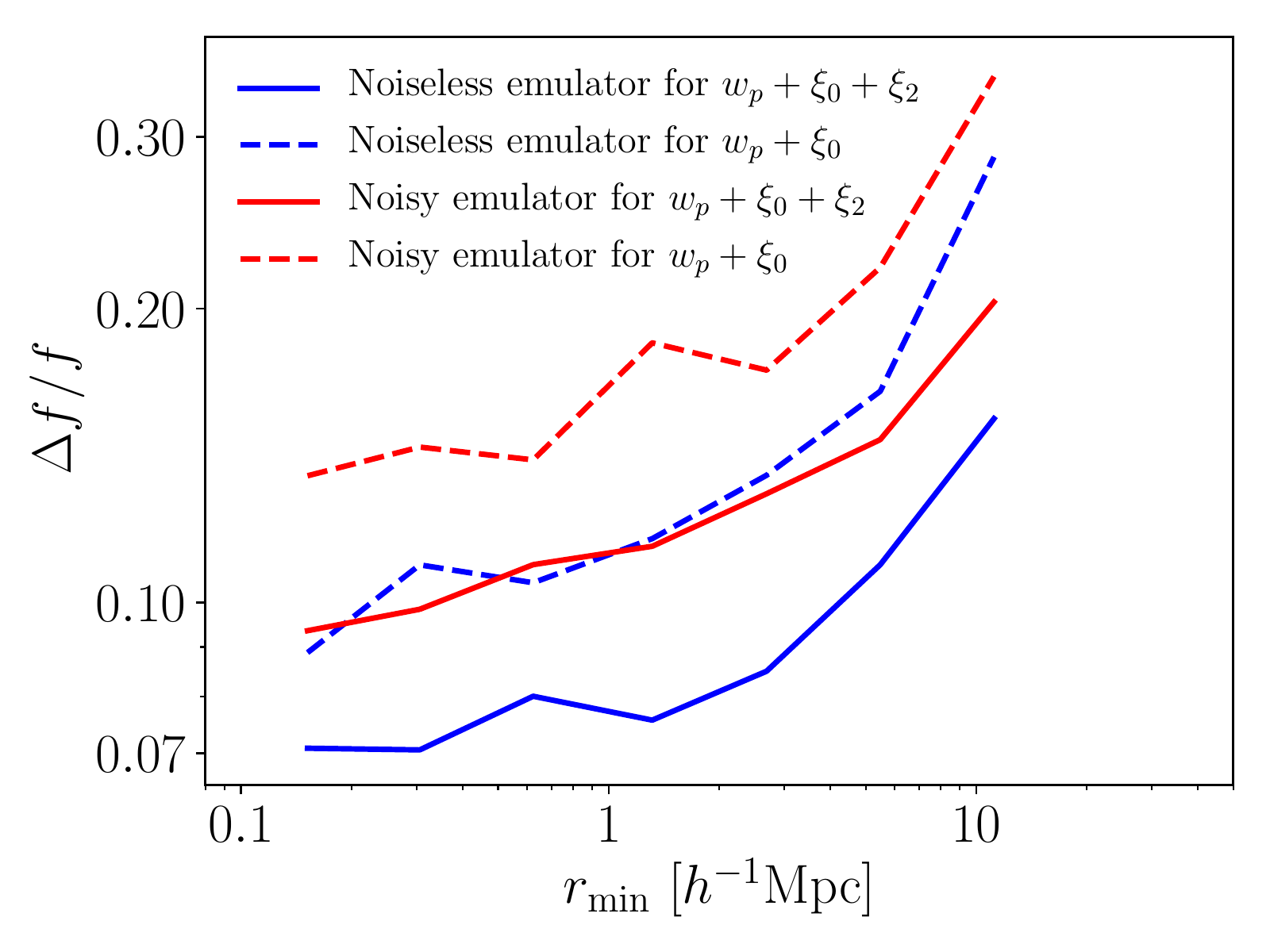}
  \includegraphics[width=0.45\textwidth]{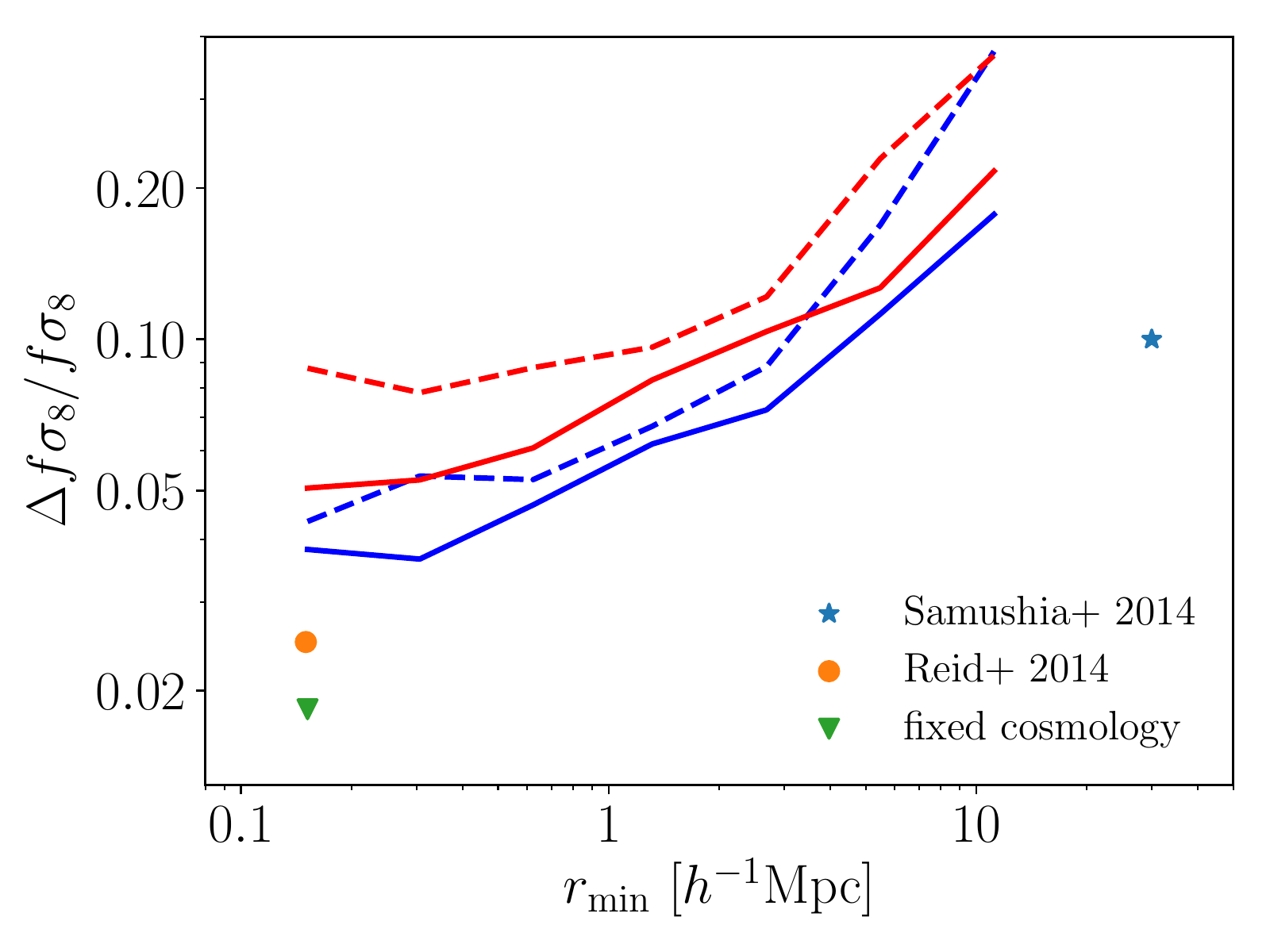}
  \caption{The scale-dependence of the constraint on $f$ (left) and $f\sigma_{8}$ (right) from galaxy clustering measurements, assuming an SDSS DR11-like galaxy sample. Solid and dashed lines correspond to constraints with and without quadrupole data respectively. Blue lines assume our emulator has no error in its prediction, i.e. the covariance matrix in \autoref{eq:likelihood} only contains sample variance, while the red lines assume the emulator has error which is added in quadrature with the sample variance. In the panel for $f\sigma_{8}$, two measurements from \cite{Reid_2014} and \cite{Samushia_2014} for BOSS CMASS galaxy sample are marked by dot and star. The triangle represents the noisy emulator test with fixed cosmological parameters. For the constraint with the noisy emulator (solid red), we find the error for $f$ and $f\sigma_{8}$ scale roughly with the minimum scale as $r_{\rm{min}}^{0.2}$ and $r_{\rm{min}}^{0.4}$ respectively.}
  \label{fig:scale_depen}
\end{figure*}

The emulator for small-scale clustering allows exploration of where the constraining power of structure growth comes from. In this test, we randomly choose a test cosmology and constrain the parameters as in Section 4.1, but now varying the minimum scale of the data from 0.1 to $10 h^{-1}\rm{Mpc}$. We perform this scale variation while also adjusting the inputs of the emulator in two ways: (1) we compare results of our standard (noisy) emulator to those where the emulator predictions are considered perfect (noiseless), and (2) we remove the quadrupole from the analysis to determine how much constraining power comes from the monopole only. The result is presented in \autoref{fig:scale_depen} for $f$ and $f\sigma_{8}$. For comparison, the results from \cite{Reid_2014} and \cite{Samushia_2014} are also shown on the panel for $f\sigma_{8}$; all emulator results have been scaled to the volume of BOSS DR11. It is clear that the cosmological information monotonically increases with decreasing minimum scale. The addition of the quadrupole tightens the constraints by a factor of 1.5 and 1.3 for $f$ and $f\sigma_{8}$ respectively. At the minimum scale, our noiseless emulator shows that a 3.6\% measurement of $f\sigma_{8}$ is possible with the DR11 volume. For the noisy emulator, the accuracy of the measurement is 5\% compared with that measured from \cite{Reid_2014} for the same survey volume. \cite{Reid_2014} employ a single simulation box which is equivalent to fixing the cosmological model. We perform a similar test with this $\delta$-function prior on the shape of the matter power spectrum and the constraint is tightened by a factor of 3 as shown by the triangle in \autoref{fig:scale_depen}. 

The above estimation can be generalized to other current or future galaxy surveys. For a DESI-like LRG survey which covers 14000 deg$^{2}$ of the sky, the probed volume from $z=0.6$ to $=1.0$ is about 12 ($h^{-1}$Gpc)$^3$, or twice the volume of one test cosmology in our simulation. This leads to a final constraint of 2$\%$ for $f\sigma_{8}$ and $4.6\%$ for $f$ respectively, in this redshift range.

\section{Discussion and Conclusion}

Using the simulations of the \aemulus Project \citep{DeRose_2018}, we have demonstrated the feasibility of constructing an emulator for the real-space and redshift-space clustering of galaxies. Using simulations independent of the training sample, our design for the emulator is able to predict the clustering of galaxies to high accuracy over a wide range of both cosmological parameter space and galaxy bias parameter space. For the scales at which current surveys yield their most precise measurements, $1<r<10$ $h^{-1}$Mpc, our model predicts the galaxy clustering signal to better than 1\%, and yields predictions that are significantly better than the sample variance of the training sample simulations at smaller and larger scales. 

The primary purpose of the galaxy clustering emulator presented here is to constrain cosmological parameters, with emphasis on the growth of dark matter structure, parameterized through $f$ and the degenerate parameter combination $f\sigma_8$. We have shown that constraints on these parameters tighten monotonically as smaller scales are included in the analysis. For a BOSS-like survey, we estimate that we can achieve 5\% accuracy on $f\sigma_8$ and 9\% accuracy on $f$ itself without using CMB priors on any other cosmological parameters. This projection for $f\sigma_8$ is two times larger than achievable through perturbation theory analysis of larger-scale information. 

As we prepare this model for application to existing data, there are several additional steps required. A fully robust model requires incorporation of galaxy assembly bias into the halo occupation model, which we will leave to a future work \citep{McLaughlin_2018}. Expanding our parameter space may degrade our constraints, thus incorporating additional observational measures may prove fruitful. Void statistics have been shown to constrain environmental dependence of halo occupation \citep{Tinker_void_2006, Tinker_void_2008}. Other statistics may also help with the degeneracies already seen in the parameter constraints in \autoref{sec:recov} and \autoref{fig:con_testBOX} (\citealt{Wibking_2017}). \cite{Guo_2015} have shown that measurements of the small-scale three-point correlation function can significantly enhance constraints on velocity bias of central and satellite galaxies, both of which show degeneracy with $f$. Among galaxy bias parameters, the strongest degeneracy with $f$ is with $M_{\rm sat}$, the mass scale of satellite galaxies. This parameter can be measured directly through galaxy clusters; indeed, the $M/N$ ratio within clusters itself contains significant cosmological information \citep{Tinker_2012,Reddick_2014}. 

In addition to reducing the theoretical uncertainties of modeling clustering at non-linear scales, the simulation-based approach used here is ideal for tackling observational systematics as well. The next version of this emulator will be applied to the existing LRG datasets, including CMASS, LOWZ and the eBOSS LRG sample at higher redshift (\citealt{Parejko_LOWZ, Zhai_2017}). The dominant observational systematic for these samples is fiber collisions---the constraint that two galaxies closer than 62 arcsec cannot be observed at the same time. Nearly all previous attempts to account for this effect involve correcting the data (e.g., \citealt{Zehavi_2011, CMASS_Martin}). By using simulations directly, it is possible to forward model the impact of fiber collisions on observational measures of clustering, and incorporate any uncertainties in the model itself. 

Although we have focused on demonstrating the constraining power of small-scale clustering for a galaxy sample of a given redshift and number density, the ultimate goal of the \aemulus galaxy clustering emulator is to build a robust tool to allow modeling of galaxies at any number density and any redshift. This will significantly increase the parameter space and the dynamic range of clustering signals to be modeled. This may require numerical algorithms beyond the traditional GP (\citealt{Ng_2014}). Our current results represent the first significant step on the path to that goal, which we expect to be a core technique in the analysis of next-generation galaxy surveys.

\acknowledgments
The authors thank Michael Blanton, Elisabeth Krause, Boris Leistedt, David Hogg, and David Weinberg for helpful comments and suggestions.  This work received support from the U.S. Department of Energy under contract number DE-AC02-76SF00515.  JLT and RHW acknowledge support of NSF grant AST-1211889. YYM is supported by the Samuel P.\ Langley PITT PACC Postdoctoral Fellowship. This research made use of computational resources at SLAC National Accelerator Laboratory, and the authors thank the SLAC computational team for support. This research used resources of the National Energy Research scientific Computing Center, a DOE Office of Science User Facility supported by the Office of Science of the U.S. Department of Energy
under Contract No. DE-AC02-05CH11231.

\software{Python,
Matplotlib \citep{matplotlib},
NumPy \citep{numpy},
SciPy \citep{scipy},
George \citep{george_2014},
Emcee \citep{ForemanMackey_2013}
}

\bibliographystyle{yahapj}
\bibliography{emu_gc_bib,software}

\end{document}